\documentclass[smallextended]{svjour3}                    
\usepackage{color}
\usepackage{amsmath,amssymb}
\usepackage{graphicx}
\usepackage{graphics}
\DeclareGraphicsExtensions{.pdf}
\usepackage{mathptmx}      
\begin{document}
	
	\title{
		Continuum approximations to systems of correlated interacting particles
	}
	
	
	\author{Leonid Berlyand$^1$         \and
	 Robert Creese$^1$    \and
		Pierre-Emmanuel Jabin$^2$     \and 
		Mykhailo Potomkin$^1$ 
	}
	
	
	\institute{
		Robert Creese \at
			\email{rec5208@psu.edu} \and  
	$^1$	Department of Mathematics, The Pennsylvania State University, University Park, Pennsylvania.  \\
	$^2$	Department of Mathematics, University of Maryland, College Park, MD 20742 USA. 
	}
	
	\date{}

	\maketitle
	
	\begin{abstract}
		We consider a system of interacting particles with random initial conditions. Continuum approximations of the system, based on truncations of the BBGKY hierarchy, are described and simulated for various initial distributions and types of interaction. Specifically, we compare the Mean Field Approximation (MFA), the Kirkwood Superposition Approximation (KSA), and a recently developed truncation of the BBGKY hierarchy (the Truncation Approximation - TA). We show that KSA and TA perform more accurately than MFA in capturing approximate distributions (histograms) obtained from Monte Carlo simulations. Furthermore, TA is more numerically stable and less computationally expensive than KSA.   
		
		\keywords{Many Particle System \and Mean Field Approximation \and Closure of BBGKY hierarchy}
	\end{abstract}
	
	\section{Introduction}
	\label{intro}
	
	
	
	\medskip 
	Systems of interacting particles are ubiquitous and can be found in many problems of physics, chemistry, biology, economics, and social science.  A wide class of such systems can be presented as follows.  
	Consider $N$ interacting particles described by a coupled system of $N$ ODEs:
	\begin{eqnarray}
	\text{d}X_i(t)= S(X_i)\text{d}t+\sqrt{2 D}\,\text{d}W_i(t)+\frac{1}{N} \sum_{j=1}^N u(X_i, X_j) \text{d}t,\quad \text{for $i=1,..,N$.} \label{IBM}
	\end{eqnarray}
	Here $X_i(t)$ is the position of the $i$th particle at time $t$, $S$ is either self-propulsion, internal frequency (as in Kuramoto model, see Subsection \ref{subsec:kuramoto}), or a conservative force field (e.g., gravity), $W_i(t)$ denotes the Weiner process, and $u(x,y)$ is an interaction force between two particles at positions $x$ and $y$. Oftentimes the force is represented by a function of the directed distance between particles, so $u$ can be written as follows
	\begin{eqnarray}\label{dist_cond}
	u(X_i, X_j)=\hat u(X_j- X_i).
	\end{eqnarray}
	
	System \eqref{IBM} has to be supplied with initial conditions. For a large number of particles $N$, finding the initial position for each particle is not practical. Instead, it is reasonable to assume that initially the positions $X_1$,$\dots$,$X_N$ are random, independent and identically distributed (i.i.d.). Thus, instead of determining a massive tuple of $N$ initial conditions, a single continuous probability distribution function is introduced.   
	
	Note that system \eqref{IBM} represents first order dynamics in which the net force is proportional to velocity, i.e. $F_i\sim V_i=\dot{X}_i$, as opposed to second order dynamics, usually obtained from Newton's Law, in which the net force is proportional to acceleration $F_i\sim a_i=\ddot{X}_i$. First order dynamics are  commonly used for models such as those of ants marching \cite{MorCapOel2005}, bacteria swimming \cite{RyaBerHai2013,RyaBerHai2011}, hierarchies in pigeons \cite{NagVásPet2013}, opinion dynamics \cite{MotTad2014}, point vortices \cite{GooHou1991}, etc.
	
	To solve \eqref{IBM} with random initial conditions means to find a joint probability distribution function (or $N$-particle pdf):
	\begin{equation*}
	f_N(t,x_1,x_2,\dots,x_N).
	\end{equation*}
	Then the probability of finding a tuple $(X_1,\dots,X_N)$ in a given domain $\Omega$ at time $t$ is 
	\begin{equation*}
	\int\limits_{\Omega}f_N(t,x_1,\dots,x_N)\,\text{d}x_1\dots\text{d}x_N.
	\end{equation*}
	The function $f_N$ can be found as a solution of the Liouville equation \cite{Spo1991}.  
	However finding a function of $N$ arguments, such as $f_N$, numerically means computing an $N$-dimen\-sional array which is prohibitively computationally expensive even for moderately large $N$. Therefore a simplification for the problem for $f_N$ is required.    
	
	A classical approach is the Mean Field Approximation (MFA) \cite{Ald1999,Jab2014,Spo1991} which relies on the assumption that initially uncorrelated particles remain uncorrelated as time evolves. 
	Then the joint probability distribution function $f_N$ is determined by a function of a two variables, $f_1(t,x)$: 
	\begin{equation}\label{mean_field_formula}
	f_N(t,x_1,x_2,...,x_N)\approx\prod\limits_{i=1}^{N}f_1(t,x_i).
	\end{equation}  
	One can substitute \eqref{mean_field_formula} into the Liouville equation for $f_N$ to obtain a partial differential equation (PDE) for $f_1$ (Vlasov equation). In the limit $N\to \infty$ (the mean field limit), formula \eqref{mean_field_formula} holds exactly \cite{BraHep1977,Dob1979}.  The function $f_1$ has the meaning of a one-particle pdf.  Alternatively, in the limit $N\to\infty$ one can describe the set of all particles as a continuum with density $f_1(t,x)$.
	Though MFA is useful in many applications, it is generally not as accurate for moderate or small $N$ \cite{MidFleGri2014}. MFA is also not applicable when the impact of correlations (which are neglected in MFA) is investigated. An example is collective behavior in bacterial suspensions \cite{SokAra2012,SokAraKes2007}: density of bacteria $-$ or equivalently one-particle pdf (since the number of bacteria is $N=10^{10}$ per $\text{cm}^3$) $-$ remains uniform, while correlation length increases, so that the two-particle pdf changes due to emergence of correlations.   
	
	
	One set of approaches to account for correlations is based on using closures of the Bogoliubov-Born-Green-Kirkwood-Yvon (BBGKY) hierarchy. This hierarchy is the system of $N$ PDEs: one  for the one-particle pdf $f_1$, one for the two-particle pdf $f_2$, ..., and one for the $N$-particle pdf $f_N$. The PDE for $f_k$ ($k=1,...,N-1$) in the BBGKY hierarchy is obtained by integration of the Liouville equation with respect to $x_{k+1},...,x_N$. The equation for $f_N$ is the Liouville equation itself. Solving the BBGKY hierarchy is equivalent to solving the Louiville equation which is computationally prohibitive as explained above. On the other hand, the PDEs in the BBGKY hierarchy are coupled as follows: the PDE for $f_k$ depends on $f_{k+1}$. Therefore, one can obtain a closed system for $f_1,...,f_k$ by introducing a closure approximation for $f_{k+1}$ in terms of $f_1,...,f_k$. For example, MFA can be considered as a closure of the BBGKY hierarchy at level $k=1$ using the closure approximation: 
	\begin{equation} \label{MF_assumption}
	f_2(t,x_1,x_2)=f_1(t,x_1)f_1(t,x_2). 
	\end{equation}
	The closure approximation \eqref{MF_assumption} means that MFA relies on the assumption that correlations in the system of interacting particles are negligible. {\it To account for correlations one needs a closure approximation at least at level $k=2$.} 
	
	The Kirkwood Superposition Approximation (KSA), developed in \cite{Kir1935}, is the most widely used closure of the BBGKY hierarchy at level $k=2$ and was applied, for example, in gas dynamics \cite{Leu1982}, simple liquids \cite{EgePagHea1971} and recently employed in biology \cite{BakSim2010,MidFleGri2014}. Following the general idea of closure approximations of the BBGKY hierarchy described above, in KSA a single ansatz for $f_3$ in terms of $f_1$ and $f_2$ is substituted in the equation for $f_2$. This ansatz is presented in Section \ref{sec:kinetic} and may be formulated in words as follows: the probability of finding the particle triple in a given configuration equals to the probability of finding each pair independently from the third particle \cite{Col1968}. Though KSA is a phenomenological ansatz, formal justification  and further improvements are available \cite{BugGorKar1991,RicLek1965}. However, we note that, up to our best knowledge,  there is no rigorous asymptotic approach to derive a closure of BBGKY hierarchy that takes into account correlations.

	Recently, a closure at level $k=2$, alternative to KSA, has been introduced in \cite{BerJabPot2016}. The main difference between KSA and the closure from \cite{BerJabPot2016} $-$ referred below to as the Truncation Approximation (TA) $-$ is that instead of a single ansatz for $f_3$, TA introduces an individual representation for each of the two terms in the equation for $f_2$ where $f_3$ appears. The choices in TA are made so that key properties of pdfs $f_1$ and $f_2$ are preserved (the properties are listed in Section \ref{sec:kinetic}). It was also proven that there is no such single representation ansatz for $f_3$ that preserves the key properties. Moreover TA is less computationally expensive than KSA.

	In this paper we consider  system \eqref{IBM} with various types of interactions $u$. We compare the closures obtained from MFA, KSA and TA with each other and with Monte Carlo Simulations of \eqref{IBM}. We show that TA is at least as accurate as KSA (when comparing to Monte Carlo
	simulations). Moreover, we observe that TA is less computationally expensive and more numerically stable than KSA. Finally for each type of interaction considered in this paper we describe the effect of correlations by comparing MFA, which neglects correlations, with other methods. Here we consider not very large $N$ for the following two reasons. First, one- and two-particle histograms $\hat{f}_1$ and $\hat{f}_2$ obtained from Monte Carlo simulations do not require excessive computations for such $N$. Note that $\hat{f}_1$ and $\hat{f}_2$ converge to true $f_1$ and $f_2$ (that is, solutions of the original not truncated BBGKY hierarchy) as sample size, the number of realizations, grows to infinity. The second reason to choose $N$ not large is to have an observable impact of correlations (correlations vanish as $N\to \infty$).            
	
	The paper is organized as follows. In Section \ref{sec:kinetic} we formulate the main problem and review and discuss the application of the BBGKY hierarchy and its truncations, such as MFA, KSA and TA. Results of numerical simulations are presented in Section \ref{sec:numerics} and discussed in Section \ref{sec:conclusion}. 
	

	\section{Description of continuum approximations}
	\label{sec:kinetic}
	In this section we review the continuum approximations that are used in this work. First, we describe the BBGKY hierarchy and then discuss its closures such as MFA, KSA, and TA. Next, we compare the three approximations noting some key differences as well as similarities between them. Numerical integration of the corresponding PDEs is presented in Section \ref{sec:numerics}.
	
	\paragraph{BBGKY Hierarchy.}
	Consider system \eqref{IBM} with $u$ satisfying \eqref{dist_cond}.  The one-particle pdf $f_1(t,x_1)$ solves the following evolution equation
	\begin{equation}\label{liouville_f_1}
	\partial_{t} f_{1}(t,x_1)+\nabla_{x_{1}}\cdot \left((S(x_1)+\mathcal{F}(t,x_1)) f_{1}(t,x_1)\right)=D\Delta_{x_1} f_1(t,x_1),
	\end{equation}
	where $\mathcal{F}$ is the conditional expectation of force exerted on the first particle which occupies the position $X_1(t)=x_1$ by all other particles:
	\begin{equation}\label{def_of_F}
	\mathcal{F}(t,x_{1})=\mathbb{E} \Bigg\{ \frac{1}{N}  \sum\limits_{j=2}^N u(X_{1}(t),X_{j}(t)) \Bigg\| X_{1}(t)=x_{1}\Bigg\}. 
	\end{equation}
	Equation \eqref{liouville_f_1} is an advection-diffusion equation for $f_1$ or, alternatively, it can be derived from the Liouville equation by direct integration with respect to all variables except $t$ and $x_1$.  
	
	An explicit formula for $\mathcal{F}$ in terms $f_1$ and $f_2$ follows from the definition of conditional expectation:
	\begin{equation}\label{edef_of_F}
	\mathcal{F}(t,x_{1})= \frac{N-1}{N}  \int u(x_{1},y) \frac{f_2(t,x_1,x_2)}{f_1(t,x_1)}\text{d}y . 
	\end{equation}
	In view of formula \eqref{edef_of_F}, we note that equation \eqref{liouville_f_1} depends on the two-particle pdf $f_2(t,x_1,x_2)$. In order to find $f_2$ we need to consider an equation for $f_2$, analogous to \eqref{liouville_f_1} for $f_1$:
	\begin{eqnarray} \label{liouville_f_2}
	\partial_{t} f_{2}(t,x_1,x_2) &+& \nabla_{x_{1}} \cdot (\mathcal{F}_{1}f_{2}(t,x_1,x_2))+\nabla_{x_{2}}\cdot (\mathcal{F}_{2}f_{2}(t,x_1,x_2)) \nonumber \\  &+& \nabla_{x_{1}} \cdot (S(x_1)f_2(t,x_1,x_2))+\nabla_{x_{2}}\cdot (S(x_2)f_2(t,x_1,x_2)) \nonumber \\ &=& D(\Delta_{x_1} f_2(t,x_1,x_2)+\Delta_{x_2} f_2(t,x_1,x_2)),
	\end{eqnarray}
	where $\mathcal{F}_{i}(t,x_1,x_2)$ ($i=1,2$) are the conditional expectation of force exerted on the $i$th particle by other particles given that $X_1(t)=x_1$ and $X_2(t)=x_2$: 
	\begin{equation} \label{eforce}
	\mathcal{F}_{i}(t,x_{1}, x_{2})=\mathbb{E} \Bigg\{ \frac{1}{N}  \sum\limits_{j\neq i}^N u(X_{i}(t),X_{j}(t)) \Bigg\| \begin{array}{c}X_{1}(t)=x_{1}\\X_{2}(t)=x_{2}\end{array}\Bigg\}.
	\end{equation}
	Using that all particles are identical and substituting conditions $X_{1}(t)=x_{1}$ and $X_{2}(t)=x_{2}$ into the sum in the right hand side of \eqref{eforce}, we simplify the formula for $\mathcal{F}_i$ 
	\begin{eqnarray} 
	\label{eforce1} \mathcal{F}_{1}(t,x_{1}, x_{2})=\frac{1}{N}u(x_1,x_2)+\frac{N-2}{N} \int\frac{  u(x_{1},y)f_{3}(t,x_{1},x_{2},y)}{  f_{2}(t,x_{1},x_{2})}\text{d}y, \\
	\label{eforce2}
	\mathcal{F}_{2}(t,x_{1}, x_{2})=\frac{1}{N}u(x_2,x_1)+\frac{N-2}{N} \int\frac{  u(x_{2},y)f_{3}(t,x_{1},x_{2},y)}{  f_{2}(t,x_{1},x_{2})}\text{d}y.
	\end{eqnarray}
	It is clear from \eqref{eforce1}-\eqref{eforce2} that to solve \eqref{liouville_f_2} one needs $f_3$, the three particle pdf.  One can write the equation for $f_3$ similar to \eqref{liouville_f_1} and \eqref{liouville_f_2}, and this equation will depend on $f_4$.  We can continue in this manner to obtain a system of $N$ coupled partial differential equations for $f_1,f_2, ... ,f_N$.  The resulting system is the  BBGKY hierarchy described in Section~\ref{intro}.  This system is prohibitively computationally expensive to solve.  Instead we look at various truncations of the BBGKY hierarchy which are computationally feasible and do not rely on the assumption that correlations are negligible unlike MFA which is a truncation in the equation for $f_1$ (at level $k=1$). Specifically, we focus on truncations in the equations for $f_1$ and $f_2$ (at level $k=2$). One can consider truncations at higher levels but the computational expense increases greatly with increasing the level of a truncation. As a result, we only consider truncations in the equations for $f_1$ and $f_2$.
	
	\medskip 
	
	\paragraph{Mean Field Approximation.} 
	MFA is a truncation of the BBGKY hierarchy at the equation for $f_1$ using the assumption
	\begin{eqnarray} \label{particles_are_uncorrelated}
	f_2(t,x_1,x_2)=f_1(t,x_1)f_1(t,x_2).
	\end{eqnarray}
	Substituting this assumption into \eqref{liouville_f_1} results in the following PDE
	\begin{eqnarray}
	\partial_t f_1(t,x_1) &+& \frac{(N-1)}{N}\nabla_{x_1} \cdot (\int u(x_1,y) f_1(t,y) \text{d}y f_1(t,x_1) ) + \nabla_{x_{1}}\cdot (S(x_1)f_1(t,x_1))  \nonumber \\ &=& D \Delta_{x_1} f_1(t,x_1).
	\end{eqnarray}
	Notice that the assumption \eqref{particles_are_uncorrelated} is equivalent to particles being uncorrelated. 
	In other words, MFA does not take into account the effects of correlations.  Taking the limit as $N \rightarrow \infty$ results in the coefficient $\frac{(N-1)}{N}$ being dropped and yields the Vlasov equation,
	\begin{eqnarray}
	\partial_t f_1(t,x_1) &+& \nabla_{x_1} \cdot (\int u(x_1,y) f_1(t,y) \text{d}y f_1(t,x_1) )+\nabla_{x_{1}}\cdot (S(x_1)f_1(t,x_1)) \nonumber \\ &=&D \Delta_{x_1} f_1(t,x_1). \label{Vlasov1}
	\end{eqnarray}
	It was shown in \cite{Dob1979} that equation \eqref{Vlasov1} is well-posed for smooth and bounded $u(x,y)$.

	The Vlasov equation \eqref{Vlasov1} can also be understood as follows. Write the BBGKY hierarchy for $N=\infty$, that is, the hierarchy is an infinite system of coupled PDEs for $f_1,f_2,\ldots$\,. Assume in addition that all particles are initially independent:  
	\begin{eqnarray}
	f_n(0,x_1, ... ,x_n)=\prod\limits_{i=1}^{n} f_1(0,x_i), \quad n\geq 1.
	\end{eqnarray}
	Then one can show that 
	\begin{eqnarray}
	f_n(t,x_1, ... ,x_n)=\prod\limits_{i=1}^{n} f_1(t,x_i) \text{ for all }t\geq 0 \text{ and } n\geq 1.
	\end{eqnarray} 
	This is so called {\it propagation of chaos}: if particles are initially independent (no correlations, ``chaotic"), then they stay independent as time evolves. 
	Propagation of chaos was shown to hold as $N \rightarrow \infty$ 
	in \cite{BraHep1977}.
	
	Therefore, MFA assumption holds exactly in the limit $N\to \infty$ and it implies that correlations in the system \eqref{IBM} are negligible.    
	Since we are interested in capturing how correlations affect the evolution of $f_1$, we must go beyond MFA. 
	
	\paragraph{Kirkwood Superposition Approximation.}
	KSA is a truncation of the BBGKY hierarchy at the equation for $f_2$. KSA is based on the following representation ansatz for $f_3$ in terms of $f_1$ and $f_2$:
	\begin{eqnarray}
	f_3(t,x_1,x_2,x_3)=\frac{f_2(t,x_1,x_2)f_2(t,x_2,x_3)f_2(t,x_1,x_3)}{f_1(t,x_1)f_1(t,x_2)f_1(t,x_3)}. \label{KSA}
	\end{eqnarray} 
	Substitute this approximation into \eqref{liouville_f_2} and obtain the following equation for $f_2$,
	\begin{eqnarray} \label{KSA_PDE}
	\partial_t f_2(t,x_1,x_2)& +&\frac{1}{N}\nabla_{x_1} \cdot \left(u(x_1,x_2)f_2(t,x_1,x_2)\right)+\frac{1}{N}\nabla_{x_2}\cdot \left(u(x_2,x_1)f_2(t,x_1,x_2)\right)\nonumber\\ 
	&+& \frac{N-2}{N}\nabla_{x_1} \cdot \int u(x_1,y) \frac{f_2(t,x_1,x_2) f_2(t,x_1,y)f_2(t,x_2,y)}{f_1(t,x_1) f_1(t,x_2)f_1(t,y)}\text{d}y\nonumber\\ &+&\frac{N-2}{N}\nabla_{x_2} \cdot \int u(x_2,y) \frac{f_2(t,x_1,x_2) f_2(t,x_1,y)f_2(t,x_2,y)}{f_1(t,x_1) f_1(t,x_2)f_1(t,y)}\text{d}y \nonumber \\ &+& \nabla_{x_{1}}\cdot (S(x_1)f_2(t,x_1,x_2)) +
	\nabla_{x_{2}} \cdot (S(x_2)f_2(t,x_1,x_2)) \nonumber \\ 
	&=& D(\Delta_{x_1} f_2(t,x_1,x_2)+\Delta_{x_2} f_2(t,x_1,x_2)).
	\end{eqnarray}
	Note that KSA representation ansatz \eqref{KSA} can be formally derived from the maximization of a truncated entropy functional \cite{Sin2004}. 
	This method can be applied to find similar approximations for $f_n$, $n>3$, however the numerical cost of solving the associated PDEs becomes prohibitive. 
	

	
	\paragraph{Truncation Approximation.}
	TA is obtained from the following observation. Consider $i=1$ and rewrite the first term in the sum \eqref{eforce} by substituting  conditions $X_1(t)=x_1$ and $X_2(t)=x_2$:
	\begin{equation}
	\label{eforce_prime}
	\mathcal{F}_1(t,x_1,x_2)=\dfrac{1}{N}u(x_1,x_2)+\mathbb{E} \Bigg\{ \frac{1}{N}  \sum\limits_{j\neq 1,2}^N u(X_{1}(t),X_{j}(t)) \Bigg\| \begin{array}{c}X_{1}(t)=x_{1}\\X_{2}(t)=x_{2}\end{array}\Bigg\}
	\end{equation} 
	Next observe that the sum in \eqref{eforce_prime} does not have a term depending on $X_2(t)$ and 
	thus it is natural to assume that the dependence of the expected value in \eqref{eforce_prime} on the condition $X_2(t)=x_2$ is weak and therefore can be ignored.  
	This observation leads to the following approximation for $\mathcal{F}_1$: 
	\begin{equation} 
	\mathcal{F}_{1}(t,x_{1}, x_{2})=\frac{1}{N}u(x_1,x_2)+\mathbb{E} \Bigg\{ \frac{1}{N}  \sum\limits_{j\neq 1,2}^N u(X_{1}(t),X_{j}(t)) \Bigg\| X_{1}(t)=x_{1}\Bigg\}. \label{TruncF1}
	\end{equation}
	A similar approximation can be written for $\mathcal{F}_2$:
	\begin{equation} 
	\mathcal{F}_{2}(t,x_{1}, x_{2})=\frac{1}{N}u(x_2,x_1)+\mathbb{E} \Bigg\{ \frac{1}{N}  \sum\limits_{j\neq 1,2}^N u(X_{2}(t),X_{j}(t)) \Bigg\| X_{2}(t)=x_{2}\Bigg\}. \label{TruncF2}
	\end{equation}
	Next we use the definition of conditional probability to rewrite \eqref{TruncF1} and \eqref{TruncF2}:
	\begin{eqnarray} \label{F_trunc1}
	\mathcal{F}_{1}(t,x_{1}, x_{2})=\frac{1}{N}u(x_1,x_2)+\frac{N-2}{N} \int  u(x_{1},y)\frac{f_2(t,x_1,y)}{ f_{1}(t,x_1)}\text{d}y, \\
	\label{F_trunc2} \mathcal{F}_{2}(t,x_{1}, x_{2})=\frac{1}{N}u(x_2,x_1)+\frac{N-2}{N} \int  u(x_{2},y)\frac{f_2(t,x_2,y)}{f_{1}(t,x_2)}\text{d}y.
	\end{eqnarray}
	Substituting \eqref{F_trunc1}-\eqref{F_trunc2} into \eqref{liouville_f_2} yields the following PDE for $f_2$, without $f_3$:
	\begin{eqnarray} \label{Trunc_PDE}
	\partial_t f_2(t,x_1,x_2)& +&\frac{1}{N}\nabla_{x_1} \cdot \left(u(x_1,x_2)f_2(t,x_1,x_2)\right)+\frac{1}{N}\nabla_{x_2} \cdot \left(u(x_2,x_1)f_2(t,x_1,x_2)\right)\nonumber\\ 
	&+& \frac{N-2}{N}\nabla_{x_1}\cdot  \int u(x_1,y) \frac{f_2(t,x_1,x_2) f_2(t,x_1,y)}{f_1(t,x_1)}\text{d}y\nonumber\\&+&\frac{N-2}{N}\nabla_{x_2} \cdot \int u(x_2,y) \frac{f_2(t,x_1,x_2 )f_2(t,x_2,y)}{ f_1(t,x_2)}\text{d}y \nonumber \\ &+& \nabla_{x_{1}}\cdot(S(x_1)f_2(t,x_1,x_2)) +
	\nabla_{x_{2}} \cdot (S(x_2)f_2(t,x_1,x_2)) \nonumber \\
	&=&D (\Delta_{x_1} f_2(t,x_1,x_2)+\Delta_{x_2} f_2(t,x_1,x_2)).
	\end{eqnarray}
	Solutions $f_1$ and $f_2$ of the system \eqref{liouville_f_1}-\eqref{Trunc_PDE} satisfy the following key properties of probability distribution functions \cite{BerJabPot2016}:
	\begin{enumerate}
		\item
		$f_2$ is symmetric with respect to $x_1$ and $x_2$: 
		\begin{eqnarray}\label{symmetry}
		f_2(t,x_1,x_2)=f_2(t,x_2,x_1).
		\end{eqnarray}
		\item
		$f_2$ conserves its mass and positivity as time evolves: 
		\begin{eqnarray}\label{mass_preservation}
		\int f_2(t,x_1,x_2) \text{d}x_1 \text{d}x_2=\int f_2(0,x_1,x_2) \text{d}x_1 \text{d}x_2, 
		\end{eqnarray}
		and
		\begin{eqnarray}
		f_1(t,x_1) \geq 0, f_2(t,x_1,x_2) \geq 0 ~\text{ if } ~ f_1(0,x_1) \geq 0, f_2(0,x_1,x_2) \geq 0 \label{positivity}.
		\end{eqnarray}
		\item  $f_1$ and $f_2$ are consistent:
		\begin{eqnarray} \label{consistency}
		f_1(t,x_1)=\int f_2(t,x_1,x_2) \text{d}x_2.
		\end{eqnarray}
		
		
		\item
		Propagation of chaos: $f_2(t,x_1,x_2)=f_1(t,x_1)f_1(t,x_2)$ where $f_1$ solves the Vlasov Equation \eqref{Vlasov1} is a solution of \eqref{Trunc_PDE} in the limit $N\to \infty$. 
		
	\end{enumerate}
	It was also shown in \cite{BerJabPot2016} that no single representation for $f_3$ is able to satisfy all four of these properties. For example, solutions of KSA, which is derived from a single representation \eqref{KSA}, do not satisfy the property of consistency \eqref{consistency}.  The fact that solutions of TA satisfy \eqref{consistency} implies that we can substitute \eqref{consistency} into \eqref{Trunc_PDE} to obtain a closed form equation for $f_2$.

	\paragraph{Comparison between approximations.}
	Here we focus on the comparison between TA and KSA since we are most interested in the effect of correlations which MFA neglects.
	First we present heuristics on how TA and KSA can be derived in a simple way. Consider a triplet of particles 1, 2, and 3 with positions at $x_1$, $x_2$, and $x_3$, respectively. 
	Assume that one studies how particles 2 and 3 affect particle 1. If correlations in the system are not low, then we need to take into account all correlations including the correlation between particles 2 and 3. On the other hand, if overall correlations are not high, then one would expect that the contribution from correlation between particles 2 and 3 only appear at a lower order for particle 1, compared to correlations between particles 1 and 2 as well as 1 and 3. Therefore, take as an approximation assumption that particles 2 and 3 are almost independent:     
	\begin{eqnarray}\label{independence_23}
	1\approx\frac{f_2(t,x_2,x_3)}{f_1(t,x_2)f_1(t,x_3)}.
	\end{eqnarray}
	Furthermore, using Bayes' Theorem and again independence of particles 2 and 3 one obtains that    
	\begin{eqnarray}
	f_3(t,x_1,x_2,x_3)&=&f_3(t,x_3 |x_1,x_2) f_2(t,x_1,x_2) \nonumber \\
	&\approx& f_2(t,x_3|x_1)f_2(t,x_1,x_2) \nonumber \\
	&=&\frac{f_2(t,x_1,x_2)f_2(t,x_1,x_3)}{f_1(t,x_1)}. \label{prob_step_1}
	\end{eqnarray}
	Here $f_3(t,x_3 |x_1,x_2)$ and $f_2(t,x_3|x_1)$ denote conditional pdfs.   
	The formula \eqref{prob_step_1} can serve as an approximation for $f_3$ with the specific assumption that  particles 2 and 3 are almost independent. To extend the formula to the case when a pair from the three particles (not specifically particles 2 and 3) is almost independent, 
	multiply \eqref{prob_step_1} by \eqref{independence_23}. By doing this we get a representation for $f_3$ which is symmetric with respect to $x_1$, $x_2$, and $x_3$,  and it exactly coincides with KSA representation \eqref{KSA}. However, multiplication by \eqref{independence_23} introduces an additional approximation error. Instead, TA uses exactly \eqref{prob_step_1} in the equation for $f_2$ where $f_3$ appears in $\mathcal{F}_1$, and \eqref{prob_step_1} with the assumption that particles 1 and 3 are almost independent in the term where $f_3$ appears in $\mathcal{F}_2$.   
	From these observations it follows that TA is more accurate than KSA, and both KSA and TA are more accurate than MFA since they are derived from less restrictive assumptions. 
	

	Finally we compare computational complexity between solving KSA and TA. To this end, note that equations for $f_2$ in KSA and TA only have a difference in the integral terms. For example, the first integral term in these equations looks as follows:
	\begin{eqnarray}
	\text{KSA:}&& \int u(x_1,y) \frac{ f_2(t,x_1,y)f_2(t,x_2,y)}{f_1(t,y)}\text{d}y\,\frac{f_2(t,x_1,x_2)}{f_1(t,x_1) f_1(t,x_2)}, \\
	\text{TA:} &&\int  u(x_{1},y){f_2(t,x_1,y)}\text{d}y\,\frac{f_{2}(t,x_{1},x_{2})}{f_1(t,x_1)}.
	\end{eqnarray}
	We see that the integral for KSA involves variables $x_1,$ $x_2$, and $y$, whereas there are only $x_1$ and $y$  for TA. This allows for the reduction of computational complexity for TA as compared to KSA since at each time step the following integral can be pre-computed: 
	\begin{equation}
	c(x):=\int  u(x,y){f_2(t,x,y)}\text{d}y,
	\end{equation}
	and used in both integral terms of the TA equation for $f_2$. Thus, TA is less computationally expensive than KSA.


	\section{Results of numerical simulations}
	\label{sec:numerics}
	
	In this section we compare numerical solutions of the continuum approximations MFA, KSA, and TA with direct simulations for various examples of interaction forces $u(X_i,X_j)=\hat{u}(X_j-X_i)$. Throughout this section, by direct simulations we mean Monte Carlo simulations of the individual based system \eqref{IBM}. First, we present our results for smooth interaction forces including positive, attraction, repulsion and attraction-repulsion interactions.  Next, we consider these continuum approximations for the Morse interaction force and the Kuramoto model. All the interaction forces are introduced below.

	In all cases, we consider dynamics of the system of interacting particles for $0<t<T$ with one-dimensional positions $X_i(t)$ and periodic boundary conditions in $0\leq x \leq 1$. For the direct simulations of system \eqref{IBM} we use the Euler-Maruyama scheme in order to capture the stochastic term. 
	The time step is $\Delta t =5\cdot 10^{-3}$ and the number of realizations is $10^6$. 
	To compute solutions for the continuum approximations a finite difference scheme was used with the spatial and time steps $\Delta x = 10^{-2}$ and $\Delta t=10^{-5}$, respectively. We note here that KSA has a specific drawback, it does not satisfy the consistency relation between $f_1$ and $f_2$, that is $f_1\neq \int f_2$. In order to find $f_1$  for KSA we use equation \eqref{liouville_f_1}. 
	The choice of the number of particles $N$ and magnitude of diffusion $D$ was made so the difference between the continuum approximations is visible and one can draw a conclusion on how the approximation captures properties of the system. For large $N$, approximations are nearly indistinguishable from each other as well as from the direct simulations, which is consistent with the mean field limit. Thus, we used small $N$, which in addition allowed us to have a reasonable computational time for the direct simulations, since they require many realizations and the computational cost of each realization depends quadratically on $N$ if \eqref{IBM} is approximated by a direct explicit method and linearly in $N$ if a more powerful particle method, such as Fast Multipole Method \cite{GreRok1987,GreRok1988,GreRok1989}, is applied.  
	We compare probability distribution functions $f_1$ and $f_2$ obtained from the continuum approximations with histograms of the positions of particles and pairs of particles obtained from the direct simulations. Here our focus is on qualitative comparison, such as description of peak formation or convergence to uniform distributions, rather than on quantitative comparison such as for example $L^p$ errors since they are not informative about the effects of correlations.

	

	\subsection{Smooth Interaction Forces} 
	\label{subsec:smooth}
	
	\indent We consider cases of positive, attracting and repulsive interaction forces as well as the one which combines short range repulsion and long range attraction. These forces are defined by \eqref{dist_cond} with $\hat{u}$ given by:
	\begin{eqnarray}
	\hat{u}_{\text{pos}}(x)&=&2\,e^{-10 x^2}, \label{def_pos}\\
	\hat{u}_{\text{att}}(x)&=&10x \,e^{-10x^2}, \label{def_att}\\
	\hat{u}_{\text{rep}}(x)&=&-10x \,e^{-10x^2}, \label{def_rep}\\
	\hat{u}_{\text{att-rep}}(x)&=&-100x(0.1^2-x^2)\, e^{-10x^2}. \label{att-rep}
	\end{eqnarray}
	
		\begin{figure}[t]  \centering
		\includegraphics[width=0.5\linewidth]{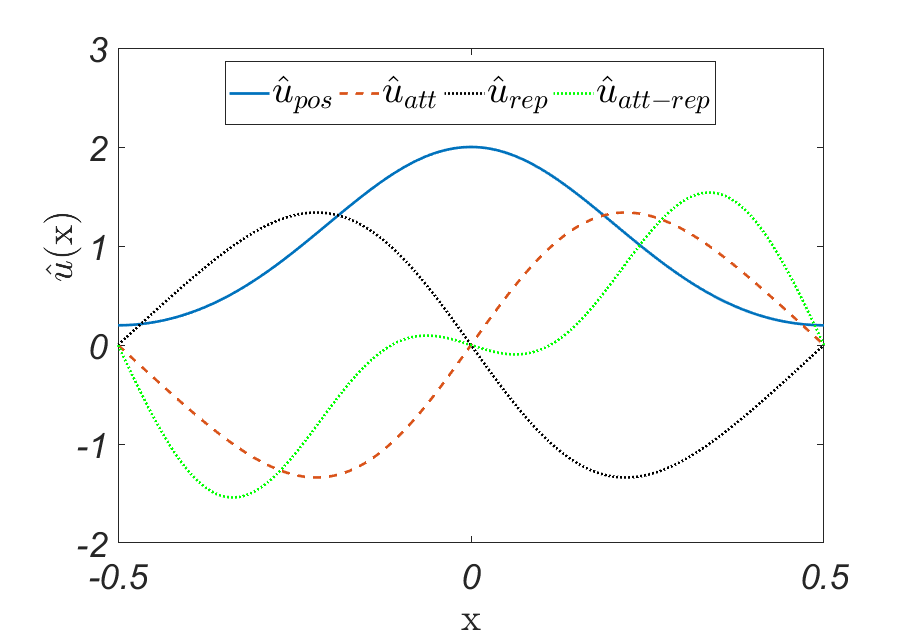}
		\caption{Plot of the smooth interaction forces given by \eqref{def_pos}-\eqref{att-rep}.  Note that the sign of $x \cdot \hat{u}(x)$ determines attraction and repulsion, with a positive sign implying attraction and a negative sign signifying repulsion. \label{interaction_plot}}
	\end{figure}

	Note that all these interaction forces are smooth functions. In particular, they are continuous at 0, unlike, for example, the Morse force, considered in the next subsection.
	Initial conditions are chosen as follows:
	\begin{eqnarray}
	\label{intcondsin}f_1(0,x)&=&1.0 + 0.4 \sin (2\pi x) \quad 0\leq x \leq 1,  \\
	\label{intcondind}f_2(0,x,y)&=&f_1(0,x)f_1(0,y) \quad 0\leq x,y\leq 1.
	\end{eqnarray}
	In other words, we consider initial one-particle distribution function $f_1$ as a perturbation of the uniform distribution $f_1\equiv 1$, and the condition \eqref{intcondind} means that the particles are initially independent. Throughout this subsection we set $N=10$ and $D=0.005$.
	
	\smallskip 
	
	\noindent{\it Positive interaction force $\hat{u}_{\text{pos}}$ given by \eqref{def_pos}.}  This force acts so that particles exert forces on each other in the positive direction only, that is, particles in front pull particles behind and those behind push those in front.  One way to think of this system is unidirectional swimming, for example fishes swimming in a narrow channel.  The fish in front will lower the resistance for those behind causing them to swim faster, while the fish behind will push the water around them forward, helping those in front to move faster. Cannibalistic locusts oriented in the same direction in a one dimensional tunnel would also follow this type of interactions, a locust will chase the locusts in front trying to eat them, while running away from those trying to eat it from behind \cite{Bazazi2008,Romanczuk2009}.
	
	 Direct simulations for this system showed that it exhibits interesting qualitative behavior for large times. Namely, particles interacting via the positive force $\hat{u}_{\text{pos}}$ tend to form a cluster which moves with speed close to $\hat{u}_{\text{pos}}(0)$. Nevertheless, the one-particle probability distribution function $f_1$ converges to a uniform value for large times, that is $f_1\approx 1$ as $t\to\infty$. This does not contradict to cluster formation, since as $t\to \infty$ the many particle system is in a highly correlated regime, therefore $f_2$ concentrates around the diagonal $x_1=x_2$ and $f_1$ is essentially the probability distribution function of the cluster location.     
	
	\begin{figure}[!htb]
		\centering
		\includegraphics[width=.45\linewidth]{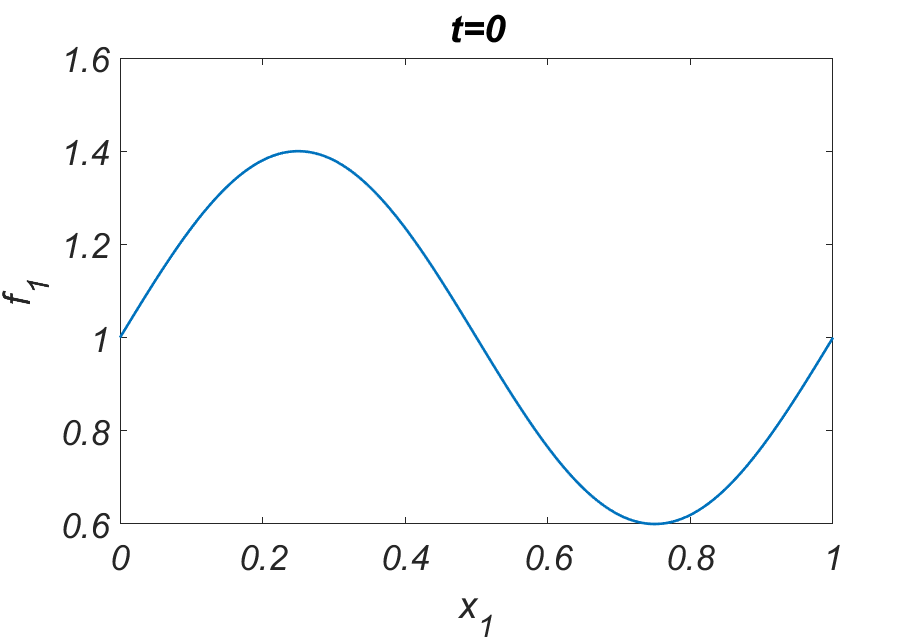}
		\includegraphics[width=.45\linewidth]{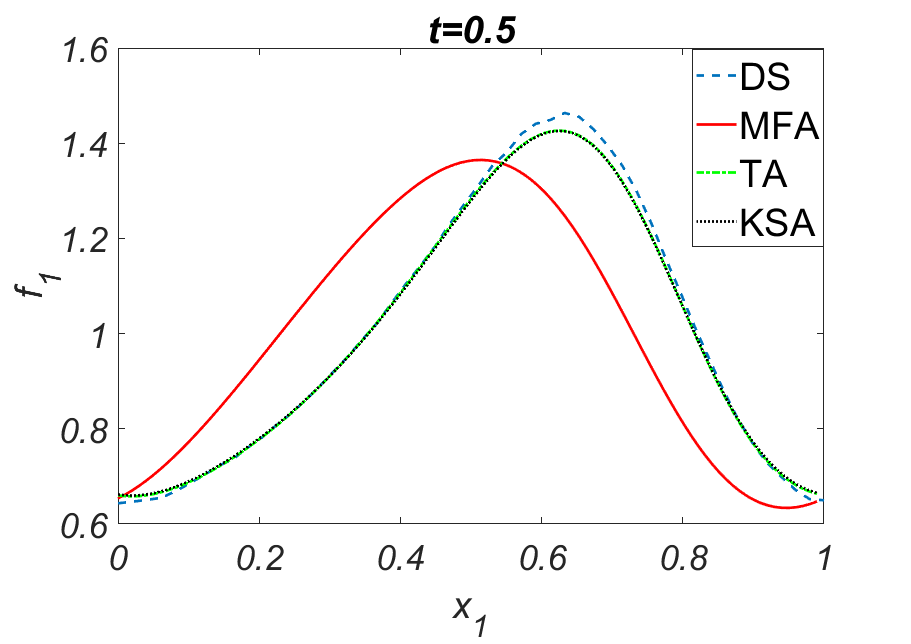}
		\caption{Left figure: the initial distribution at $t=0$ for all approximations.  Right figure: $f_1(0.5,x)$ for the various approximations to \eqref{IBM} with the positive interaction force $\hat{u}_{\text{pos}}(x)=e^{-12 x^2}$ with $N=10$ and initial conditions given by \eqref{intcondsin}.}   \label{Pos1fig}
	\end{figure}

	In direct simulations we observe that the peak of $f_1(t,x)$ moves to the right and slightly grows as time increases. Motion to the right is because all the particles' velocities in this case are positive, that is, $\dot{X}_i(t)>0$ (if diffusion is disregarded).  The growth of the peak is due to the particles clustering.
	
\begin{figure}[!htb]
		\begin{center}
			\includegraphics[width=0.45\textwidth]{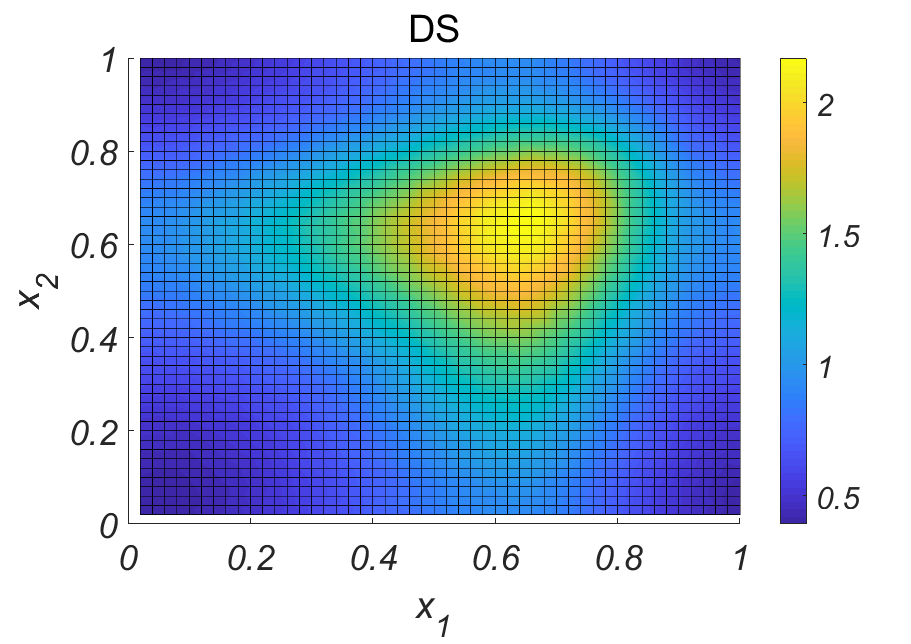}
			\includegraphics[width=0.45\textwidth]{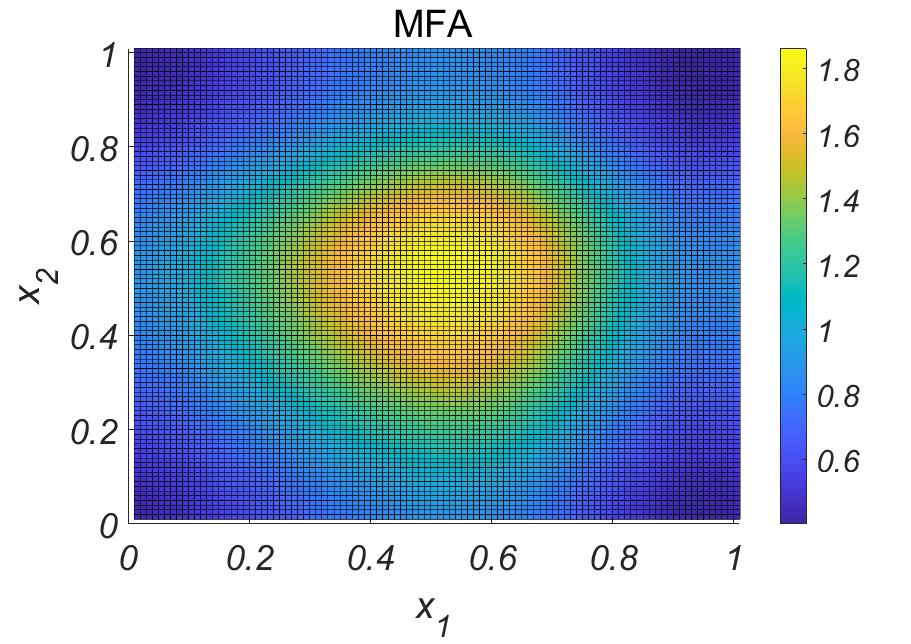}\\
			\includegraphics[width=0.45\textwidth]{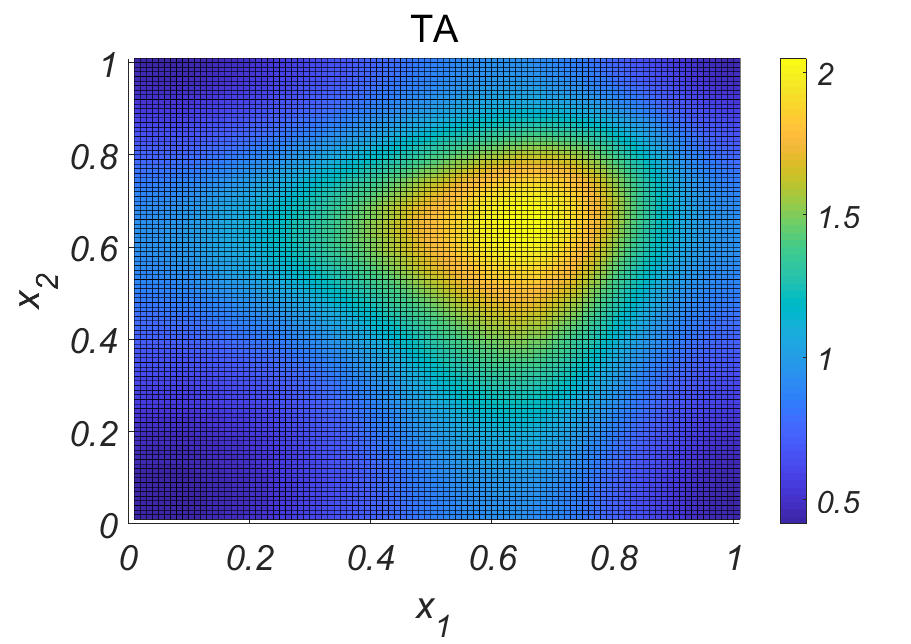}
			\includegraphics[width=0.45\textwidth]{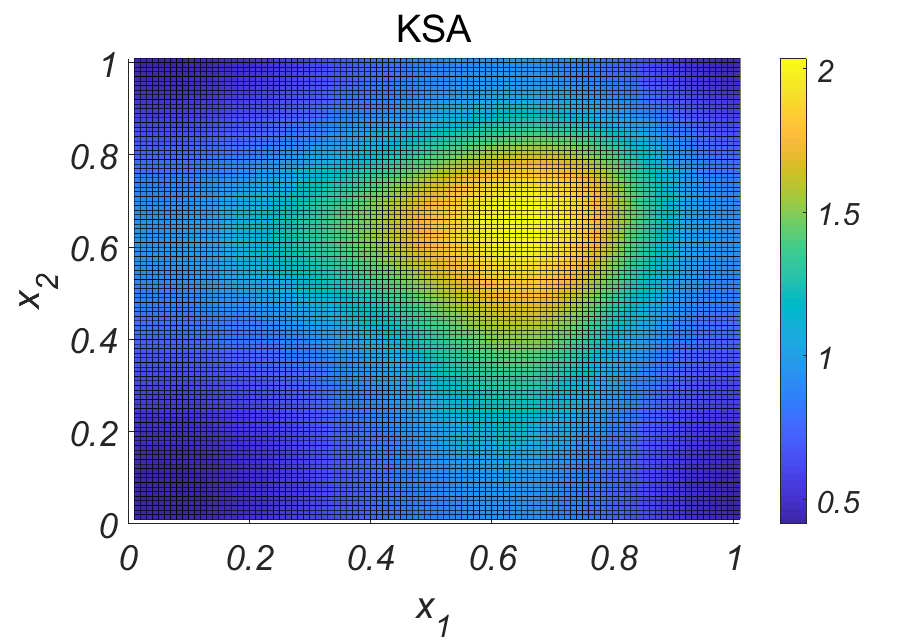}
			\caption{The figures show the approximations for $f_2$ at $t=0.5$ with the positive interaction force as follows:  top left: Direct Simulations,  top right: Mean Field, bottom left: Truncation Approximation, bottom right: Kirkwood Superposition Approximation}
			\label{posfigf2}
		\end{center}	
	\end{figure}

	In continuum approximations the solution $f_1$ for both KSA and TA moves to the right at almost the same pace as for the direct simulations and also captures the growth of the peak while MFA moves slower and is unable to capture the growth of the peak, see fig. \ref{Pos1fig}.  This is due to particles clustering and that particles move faster when they are part of a cluster.
	Since these effects come from correlations, the methods such as TA and KSA, which take into account correlations, capture the speed and the peak growth better than the MFA.  In fig. \ref{posfigf2}, one can see that like in the direct simulations the two-particle distribution function $f_2$ computed by TA and KSA has a single non-round peak (the yellow spot), while $f_2$ in MFA has smaller wider and round peak. 

	\smallskip

\noindent{\it Attracting interaction force $\hat u_{\text{att}}$ given by \eqref{def_att}}.
	This force results in particles approaching to one another and as time evolves the particles tend to concentrate at a single location determined by initial conditions. As in the case of positive interaction force, for $N<\infty$ one should distinguish between the concentration of many interacting particles and one-particle probability distribution function $f_1$. While the particles tend to cluster at a single location, the one-particle probability distribution function does not become a $\delta$-function. Moreover, if initially the distribution $f_1$ is close to uniform, it stays nearly uniform for all $t>0$, even though the particles will almost surely form a point cluster. This is because particles tend to concentrate but the point of concentration is random and almost uniformly distributed. On the other hand, for fixed initial $f_1$, if $N$ increases, then the one-particle distribution function $f_1$ eventually (i.e., as $t\to \infty$) exhibits larger peaks (unless it is initially uniform), and in the limit $N\to \infty$ it becomes a $\delta$-function. This is consistent with the mean field limit: as $N\to\infty$ correlations vanish, and the notion of one-particle probability distribution function $f_1$ coincides with the particles concentration. We also note that the two-particle distribution function $f_2(t,x_1,x_2)$ for all $N$ concentrates along the diagonal $x_1=x_2$ as $t\to \infty$.

	\begin{figure}[!htb]
		\begin{center}
			\includegraphics[width=.45\textwidth]{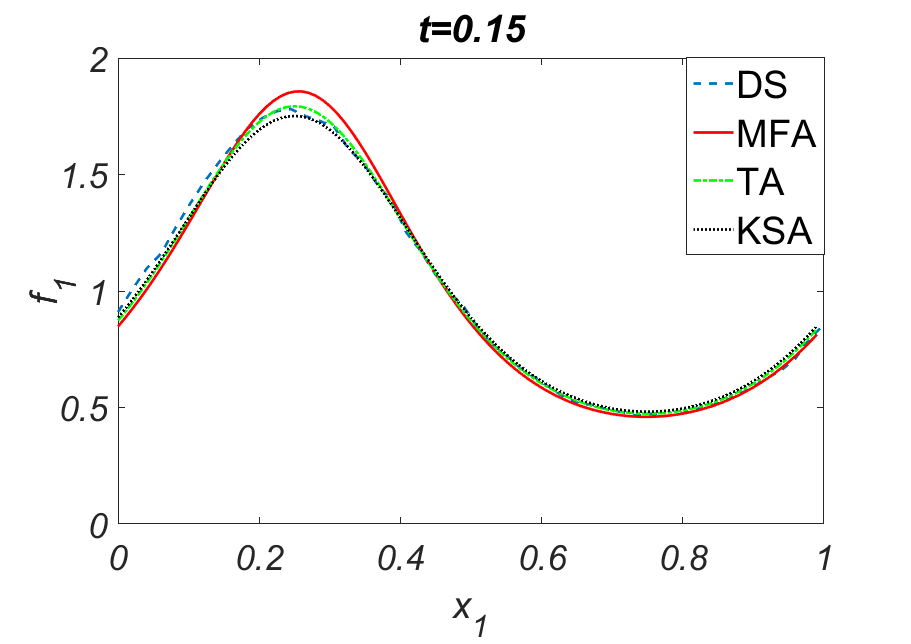} 
			\includegraphics[width=.45\textwidth]{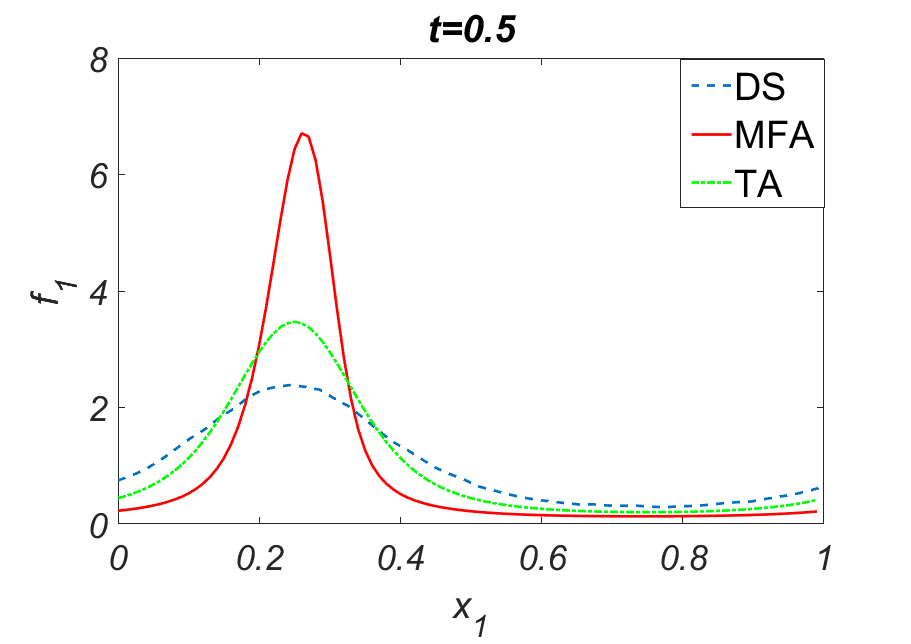}
			\caption{Approximations of $f_1$ for $\hat u_{\text{att}}(x)=10x\,e^{-10x^2}$ with $N=10$ and initial conditions given by \eqref{intcondsin}. Left: $t=0.15$, right: $t=0.5$.}\label{Att1fig}
		\end{center}
	\end{figure}
	 
\begin{figure}[!htb]
		\begin{center}
			\includegraphics[width=.45\textwidth]{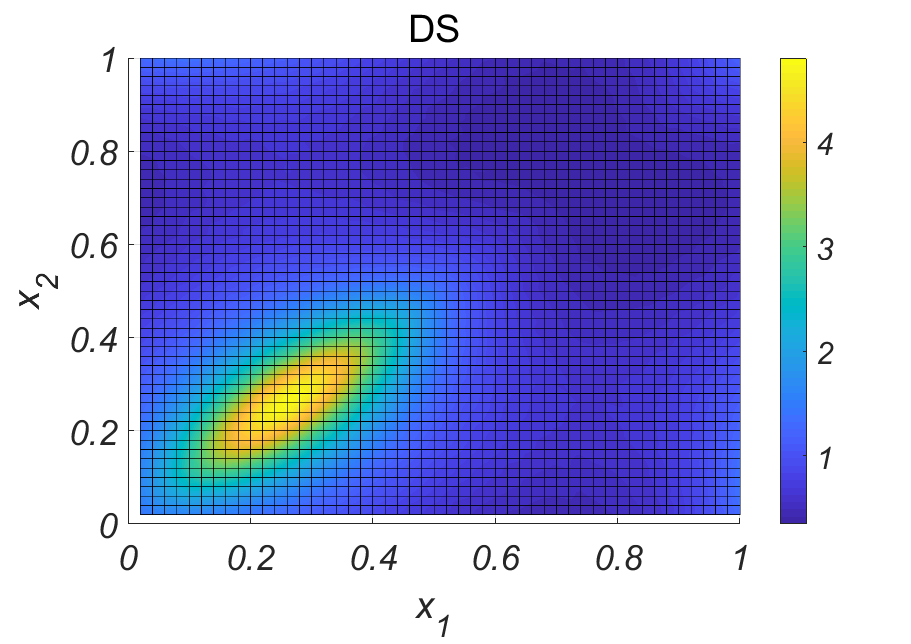}
			\includegraphics[width=.45\textwidth]{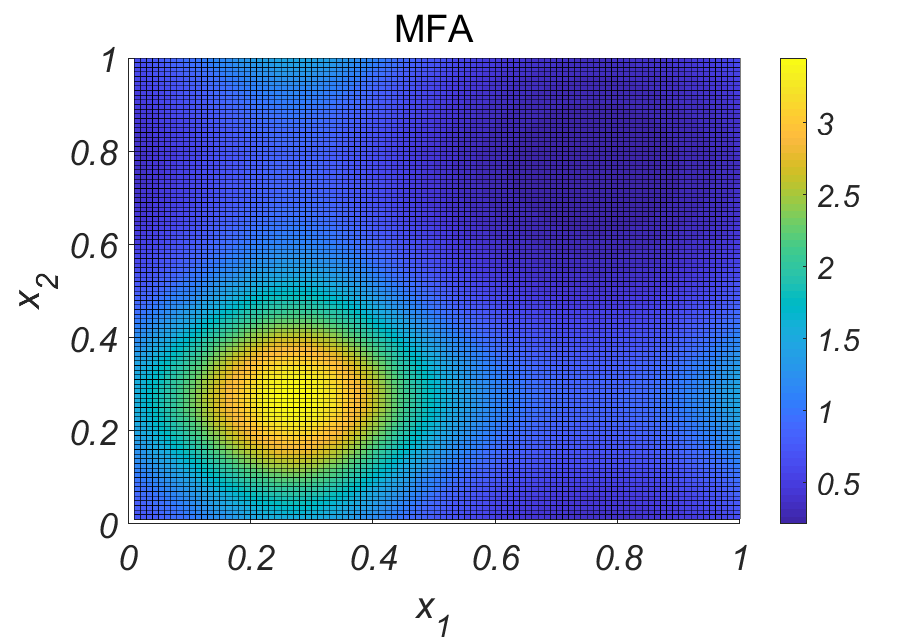}
			\\
			\includegraphics[width=.45\textwidth]{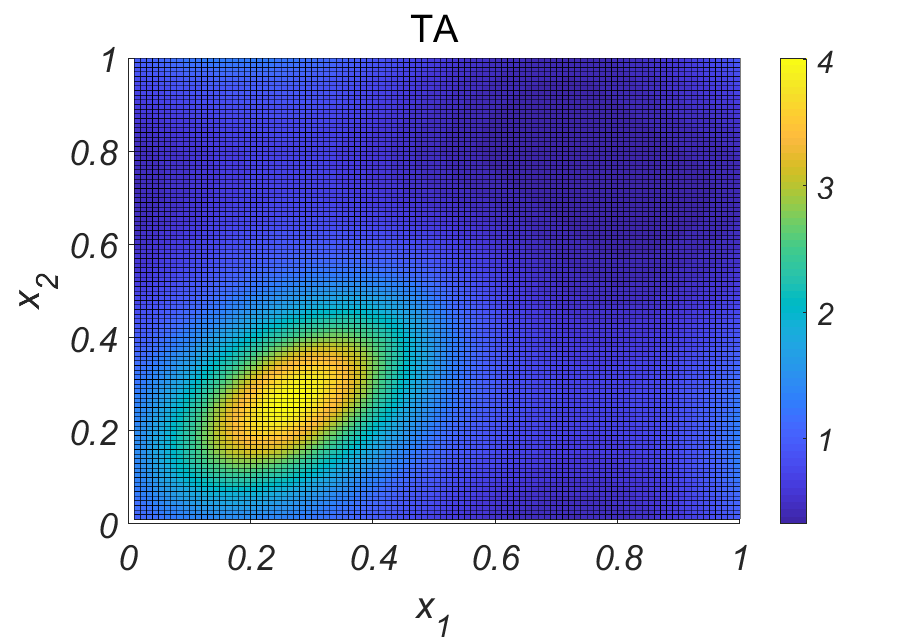}
			\includegraphics[width=.45\textwidth]{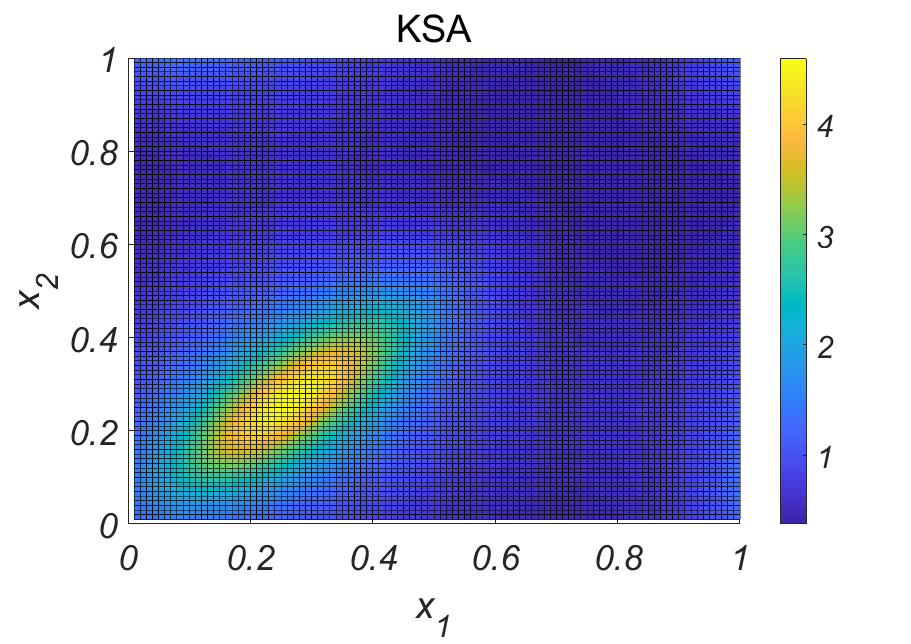}
			\caption{Approximations for $f_2$ at $t=0.15$ with the attraction interaction force as follows.  Top left: Direct Simulations, top right: Mean Field Approximation, bottom left: Truncation Approximation, bottom right: Kirkwood Superposition Approximation. }	\label{Attfigf21}
		\end{center}
	\end{figure}

	All three approximations capture the growth of the peak of $f_1$  at $t=0.15$, see fig. \ref{Att1fig} (left).  The MFA overestimates the peak but KSA and TA both capture it well with TA being slightly more accurate.  Both KSA and TA, unlike MFA, capture large values of $f_2$ near the diagonal, $x_1=x_2$ (see fig.~\ref{Attfigf21}: the peak represented by the yellow spot is elongated along the diagonal for KSA and TA, while for MFA it is round). Recall that concentration of $f_2$ near the diagonal $x_1=x_2$ means that any two particles are located close to each other. 
	All approximations underestimate the maximum value of $f_2$ obtained from the direct simulations, with KSA being the closest.  
	\begin{figure}[!htb] 
	\begin{center}
		\includegraphics[width=.45\textwidth]{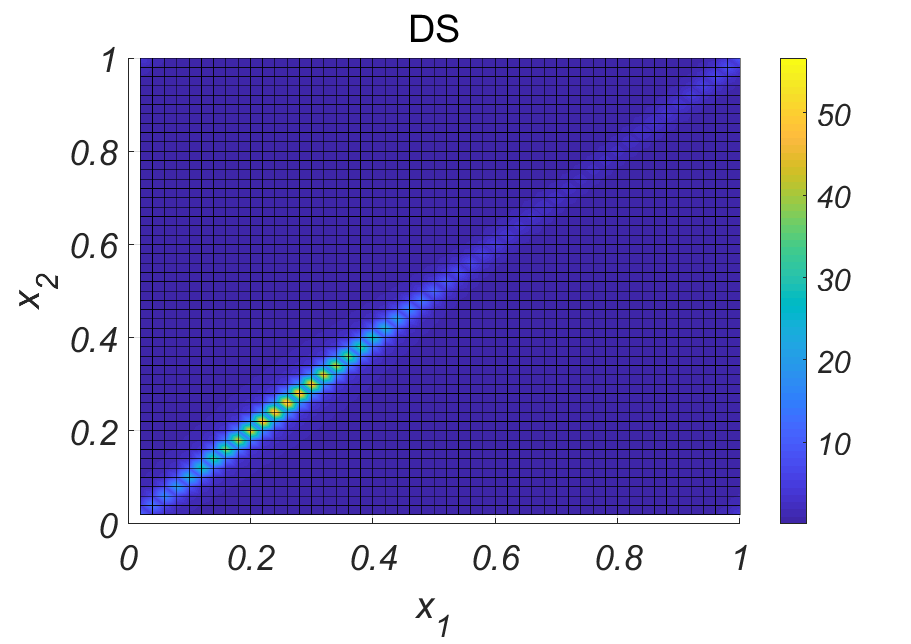}
		\label{Att2figf21}	\\
		\includegraphics[width=.45\textwidth]{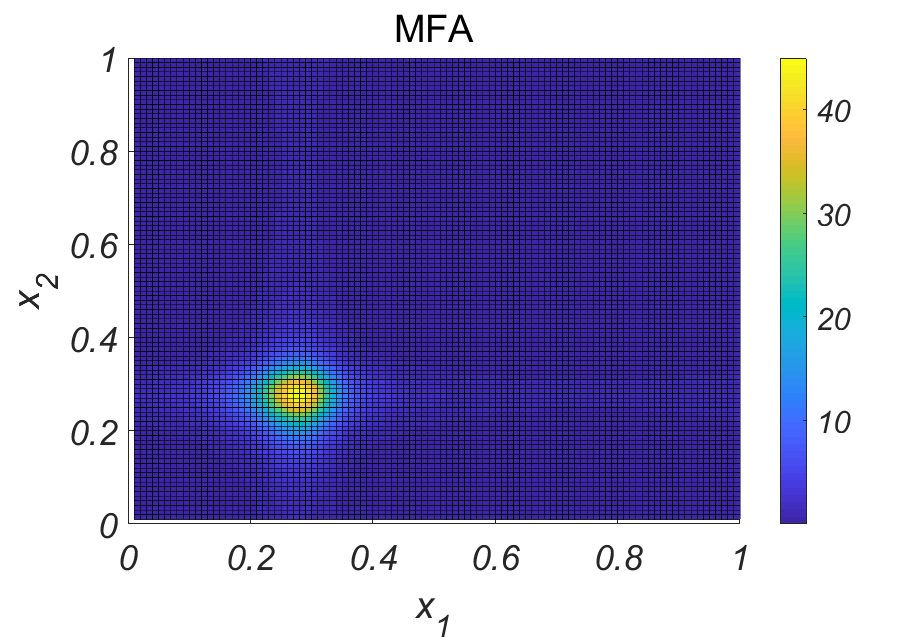}
		\includegraphics[width=.45\textwidth]{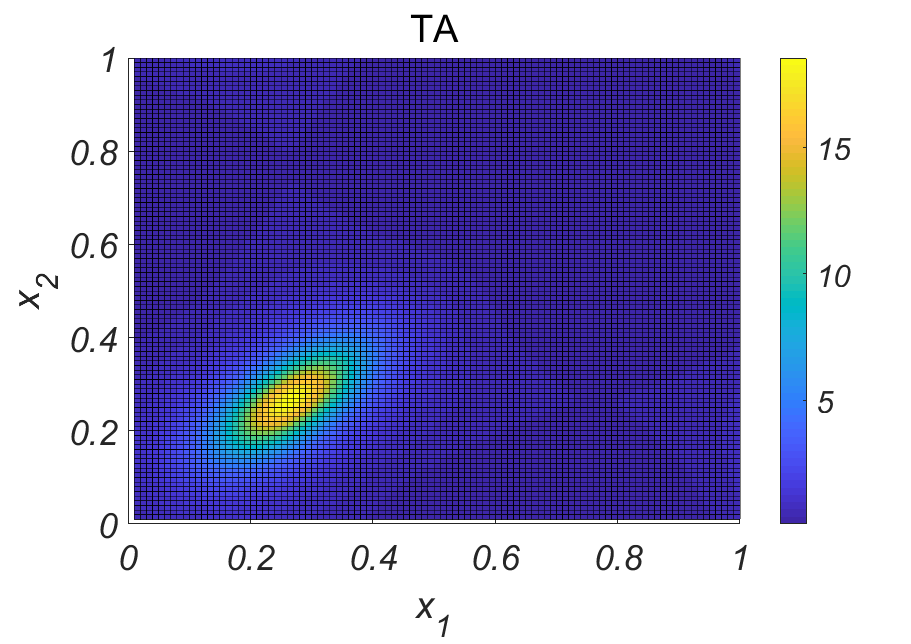}
		\caption{Approximations for $f_2$ at $t=0.5$ with the attraction interaction force as follows. Top: Direct Simulations, bottom left: Mean Field, bottom right: Truncation Approximation. } \label{Attfigf22}
	\end{center}
\end{figure}
	At $t=0.5$  MFA greatly overestimates the growth of the peak of $f_1$, TA also overestimates the growth of the peak but is much closer to the direct simulations than MFA, see fig.~\ref{Att1fig} (right). Moreover, among the two truncations at level $k=2$ considered in this work, TA was capable of producing results with the explicit numerical scheme, while the numerical simulations for KSA became unstable and are not presented. For $f_2$ TA approximates values obtained from direct simulations near the diagonal better than MFA,  see fig. \ref{Attfigf22}.  
	The maximal value of $f_2$ in direct simulations is also closer to TA than MFA. 

	\noindent{\it Repulsion interaction force $\hat u_{\text{rep}}$ given by \eqref{def_rep}.} This force results in particles pushing away from one another.  As time evolves, particles form a lattice with even spacing, where the final locations are determined by initial conditions. The one-particle probability distribution function $f_1$ becomes uniform as $t\to\infty$. Repulsion between particles leads to that values of the two-particle probability distribution function $f_2$ near the diagonal $x_1=x_2$ decrease with time, and $f_2$ concentrates at lines $|x_1-x_2|=\frac{k}{N}$, $k=1,...,N$ as $t\to \infty$ (these lines are parallel to the diagonal $x_1=x_2$ but the diagonal is not one of these lines). 
	
\begin{figure}
		\begin{center}
			\includegraphics[width=.5\textwidth]{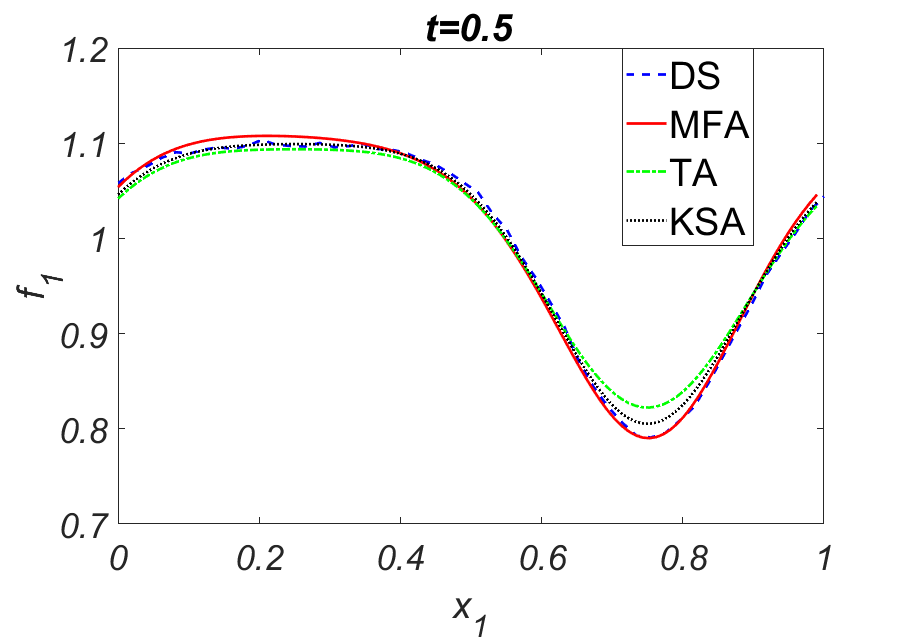}
			\caption{Approximations of $f_1(0.5,x_1)$ for the repulsion interaction force $\hat u_{\text{rep}}$ with $N=10$ where initial conditions are given by \eqref{intcondsin}.}
				\label{repfig}
		\end{center}
	\end{figure}

\begin{figure}[!htb] \begin{center}
			\includegraphics[width=.45\textwidth]{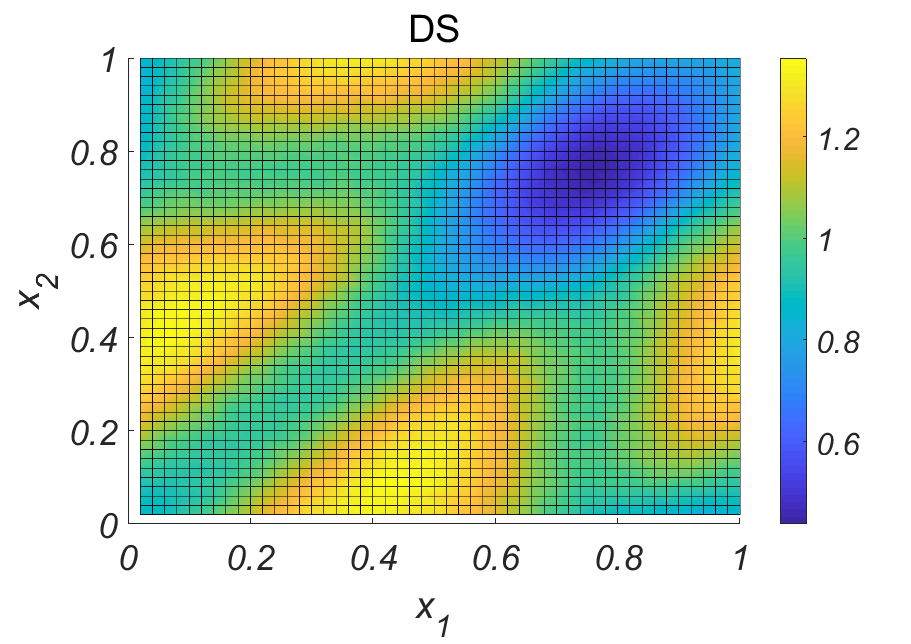}
			\includegraphics[width=.45\textwidth]{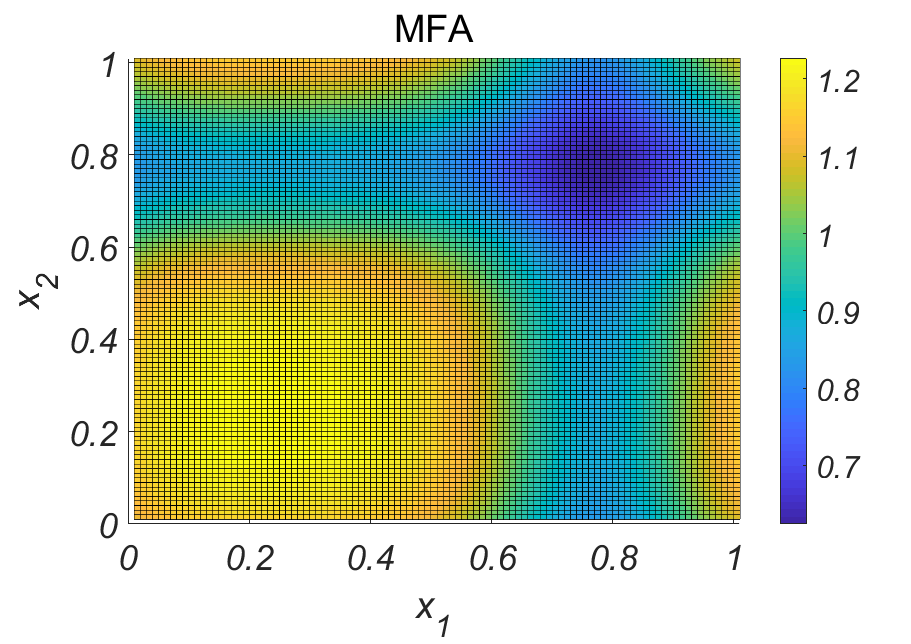} \\
			\includegraphics[width=.45\textwidth]{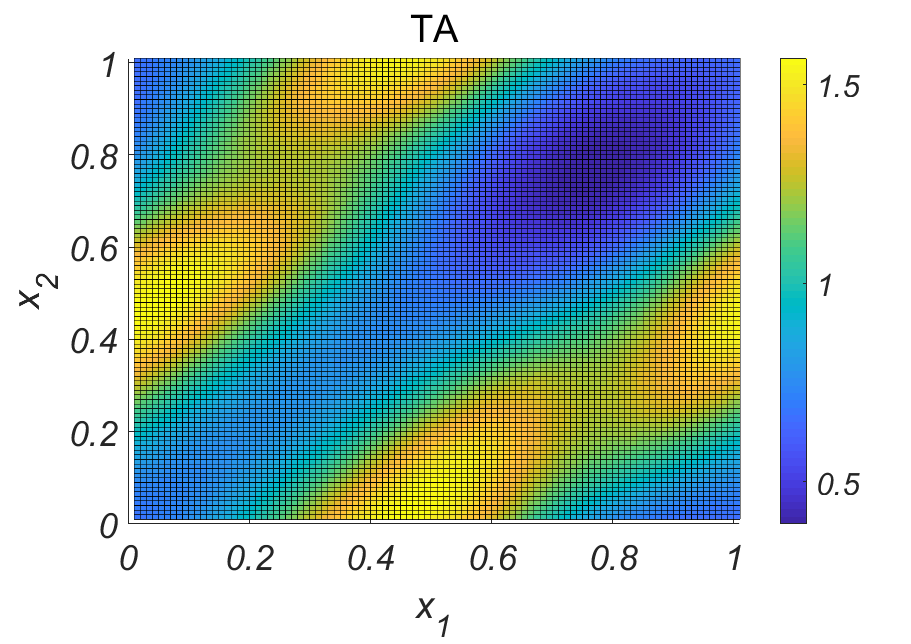}
			\includegraphics[width=.45\textwidth]{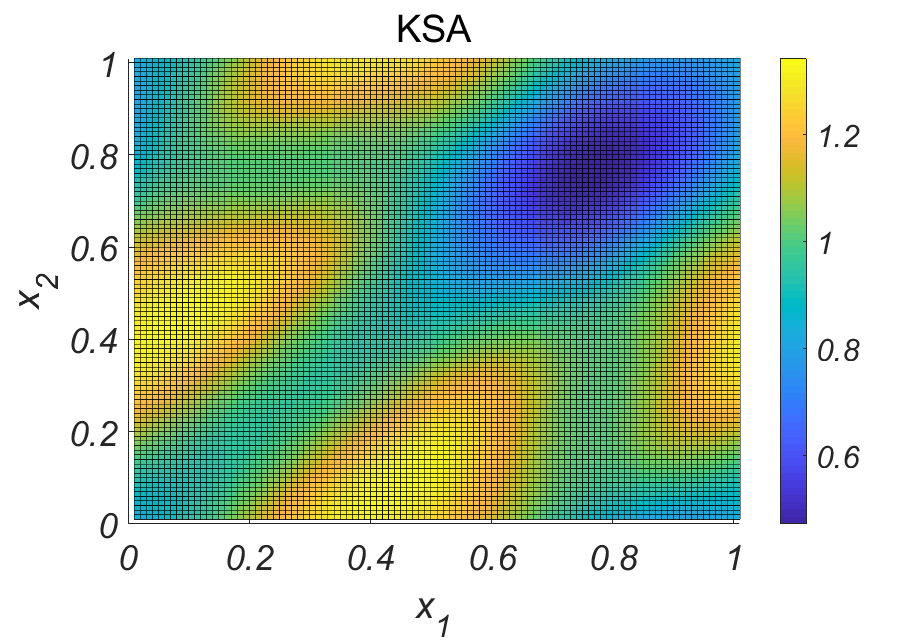}
			\caption{Approximations of $f_2$ at $t=0.5$ for $\hat{u}_{\text{rep}}$.  Top left: Direct Simulations, top right: Mean Field Approximation, bottom left: Truncation Approximation, bottom right: Kirkwood Superposition Approximation. }
			\label{Repfigf21}
		\end{center}
	\end{figure}

		All three approximations capture tendency of $f_1$ to become uniform, see fig.~\ref{repfig}. 
	The KSA and the TA, unlike the MFA, capture that the values of $f_2$ along the diagonal $x_1=x_2$ decrease in time, see fig.~\ref{Repfigf21}.


	
	\smallskip
	
	\noindent{\it Interaction force with repulsion at short range and attraction at long range $\hat u_{\text{att-rep}}$ given by \eqref{att-rep}}.
	This interaction force results in particles pushing away from one another when the distance between the particles is less than $0.1$ and attracting otherwise. 
	Interaction forces which are repulsive at short range and attracting at long range are very common in physics. For example, in order to describe forces between atoms, a variety of such interaction functions introduced via potentials is used, among them the Morse, the Yukawa, and the Lennard-Jones potentials. The Morse potential will be considered in Section~\ref{subsec:morse}. The main difference between $\hat u_{\text{att-rep}}$ and these potential forces that $\hat u_{\text{att-rep}}$ is smooth at $0$ which leads to that the repulsion part of $\hat u_{\text{att-rep}}$ is weaker than for the potential forces and hence it can not be considered as a good choice for modeling if interactions between particles at short range are steric, that is, particles have a finite size and do not penetrate each other. However, since equations for $f_1$ and $f_2$ contain derivatives of terms with $\hat{u}$, a smooth interaction forces are the most convenient for numerical simulations among all short-range-repelling/long-range-attracting interaction forces.  
	    
	\begin{figure}
		\begin{center}
			\includegraphics[width=.5\linewidth]{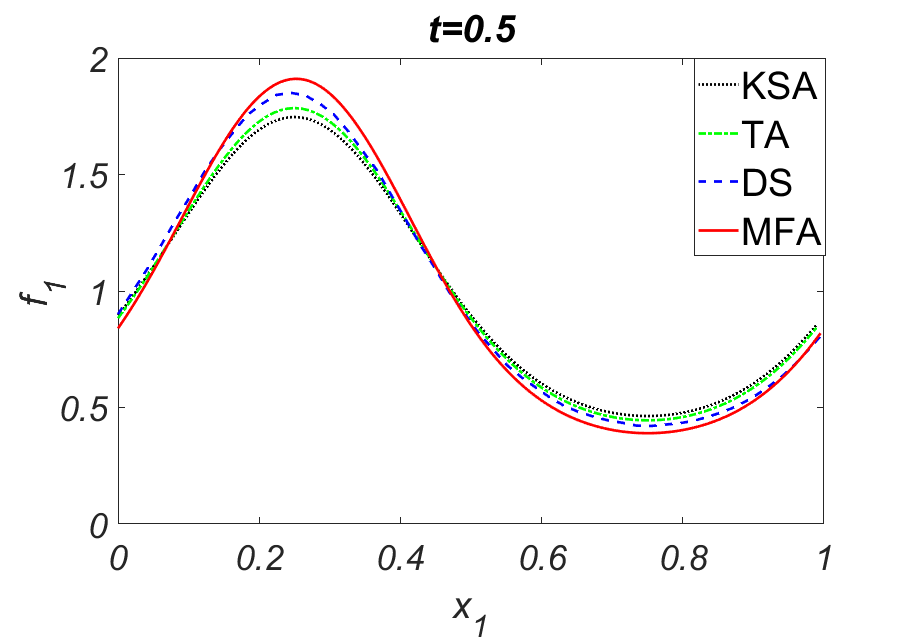}
			\caption{Approximations of $f_1(0.5,x)$ for $\hat u_{\text{att-rep}}$ with $N=10$ where initial conditions are given by \eqref{intcondsin}.}
			\label{smooth3fig}
		\end{center}
	\end{figure}
	\begin{figure}[!htb]
		\begin{center}
			\includegraphics[width=.45\textwidth]{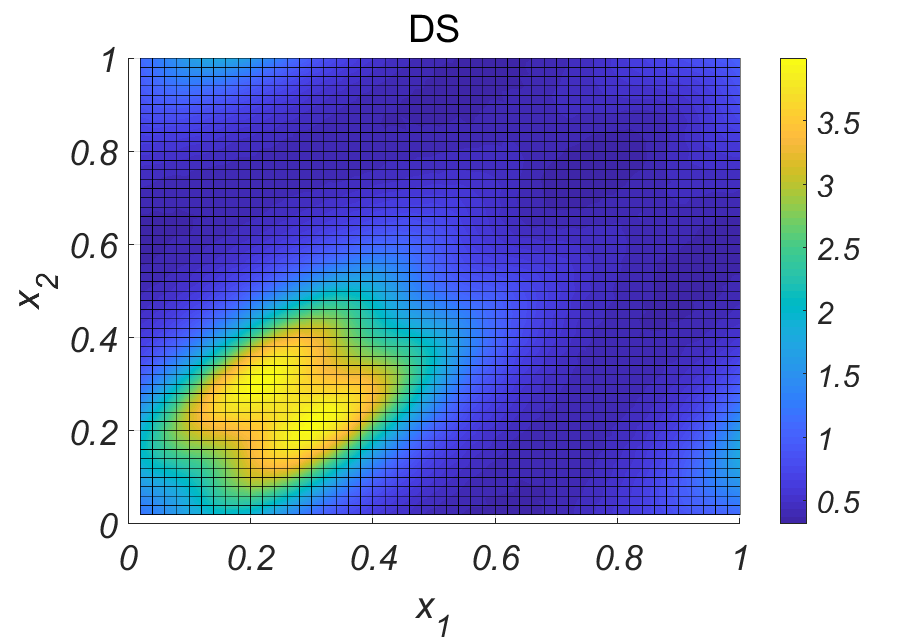}
			\includegraphics[width=.45\textwidth]{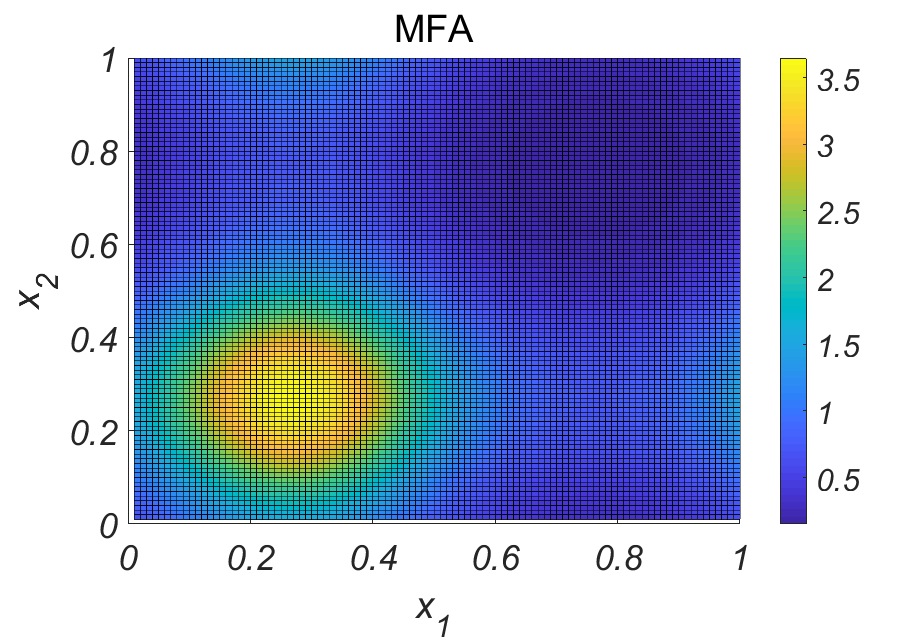}\\
			\includegraphics[width=.45\textwidth]{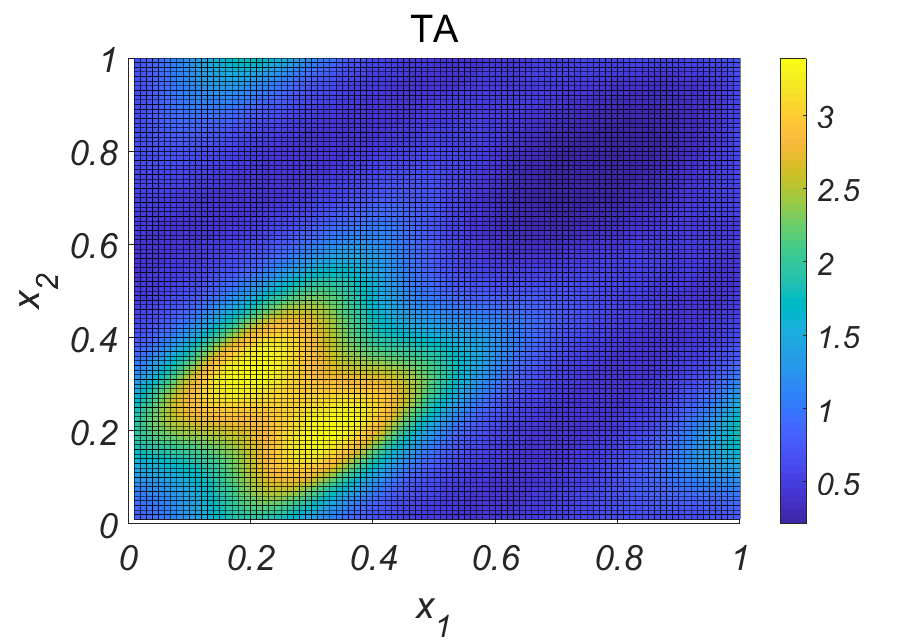}
			\includegraphics[width=.45\textwidth]{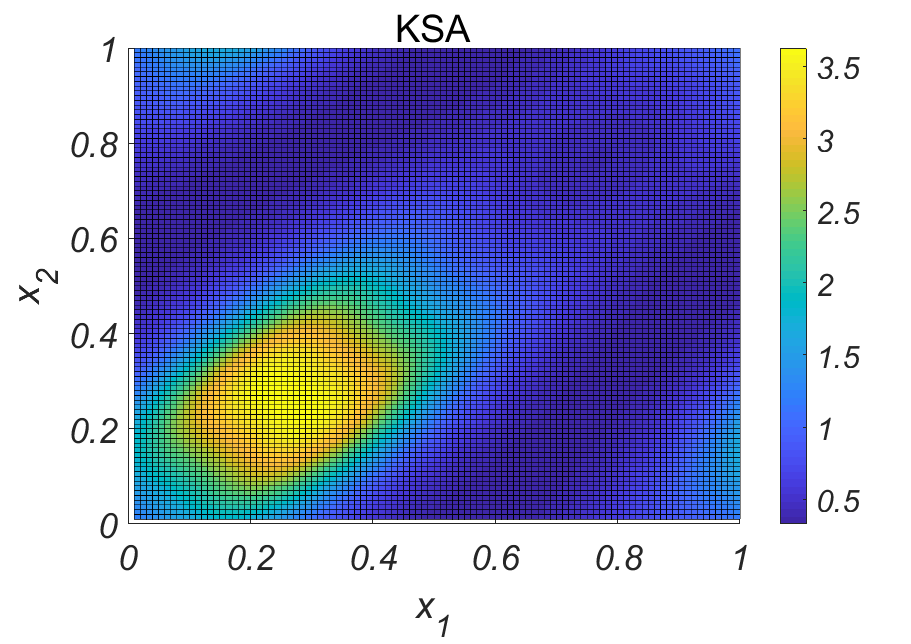}
			\caption{Approximations of $f_2$ at $t=0.5$ for $\hat{u}_{\text{att-rep}}$. Top left: Direct Simulations, top right: Mean Field, bottom left: Truncation Approximation, bottom right: Kirkwood Superposition Approximation. }
			\label{S3figf21}
		\end{center}
	\end{figure}
	
	Results of numerical simulations for all approximations of one-particle probability distribution function $f_1$  with interaction force $\hat{u}_{\text{att-rep}}$ are depicted in fig.~\ref{smooth3fig}. MFA overestimates the peak of $f_1$ obtained from direct simulations. Note that this observation is similar to the one for the attraction interaction force, see fig.~\ref{Att1fig}; this is because the attraction component of interactions dominates due to the specific form of $\hat{u}_{\text{att-rep}}$, see fig.~\ref{interaction_plot}.  However, TA and KSA underestimate the peak; this is presumably because these approximations overestimate the repulsion at the peak as in fig.~\ref{repfig}. TA, unlike MFA and KSA, captures that $f_2$ tends to decrease at the diagonal $x_1=x_2$.    
	
	

	\subsection{Morse interaction force}
	\label{subsec:morse}
	The Morse interaction force was originally introduced in physics and chemistry to model inter-atomic forces, see e.g. \cite{Morse1929,Schiff1968}, and was further used in other disciplines such as for example mathematical biology, see \cite{Newman2005,MidFleGri2014}. As in the case of $u_{\text{att-rep}}$ from Subsection~\ref{subsec:smooth}, particles which interact through the Morse interaction force repel each other if they are close and attract otherwise. On the other hand, unlike $u_{\text{att-rep}}$, the repulsion of the Morse interaction force does not vanish as particles approach each other. The growth of repulsion as inter-particle distance goes to zero is relevant if for instance the repulsion component serves to model flexible volume constraints (that is, particles push each other away if the share the same place). 
	Specifically, the system of many particles interacting through the Morse interaction force is    
	\begin{eqnarray} \label{Morse1}
	\text{d}X_i&=&\sum_{j \neq i} \hat u(X_j-X_i)\,\text{d}t +\sqrt{2D} \,\text{d}W_t, \text{ where} \\ 
	&&~~\hat u_{M}(x)=\left\{\begin{array}{ll}120\left[e^{-2(|x|-r_e)}-e^{-(|x|-r_e)}\right]\dfrac{x}{|x|},  & |x| \leq c, \\ 0,&|x|>c. \end{array} \right.\label{Morse2}
	\end{eqnarray}
	
\begin{figure}[!htb]  \centering
		\includegraphics[width=\linewidth]{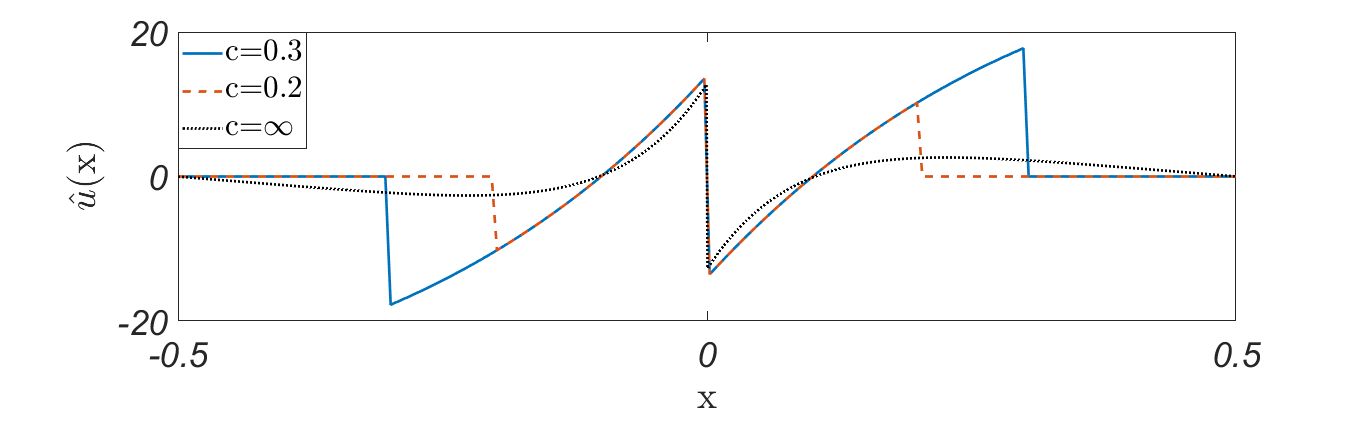}
		\caption{The Morse interaction force with truncations at $c=0.2$ and $c=0.3$.  Note that the sign of $x \cdot \hat{u}(x)$ determines attraction and repulsion, with a positive sign implying attraction and a negative sign signifying repulsion. Dashed line depicts the plot for the Morse force for $c=\infty$ (rescaled for better visibility).}
		\label{interaction_plot1}
	\end{figure}

	The Morse interaction force is defined in \eqref{Morse2}, see fig.~\ref{interaction_plot1}. The parameter $r_e=0.1$ is the equilibrium distance, that is, the distance at which the force vanishes, and $c$ is the radius of truncation of the Morse force or in other words $c$ is the range of interactions.  
	
	For the numerical simulations of system \eqref{Morse1}-\eqref{Morse2} the following initial conditions were chosen 	
	\begin{equation}
	 f_1(t=0,x)=\dfrac{5}{6} (\tanh(30(x-0.2))+\tanh(30(0.8-x))).
	 \label{tanIntCond}
	\end{equation}
	We note here that we chose to present results of numerical simulations for initial conditions \eqref{tanIntCond} instead of \eqref{intcondsin} from Subsection \ref{subsec:smooth}, since the system \eqref{Morse1}-\eqref{Morse2} for latter conditions showed very slow dynamics of the probability distribution function $f_1$.  Visible changes in $f_1$ with initial condition \eqref{tanIntCond} as time evolves are due to large gradients of $f_1$ at $x\approx 0.2$ and $x\approx 0.8$.

\begin{figure}
		\begin{center}	
			\includegraphics[width=.5\textwidth]{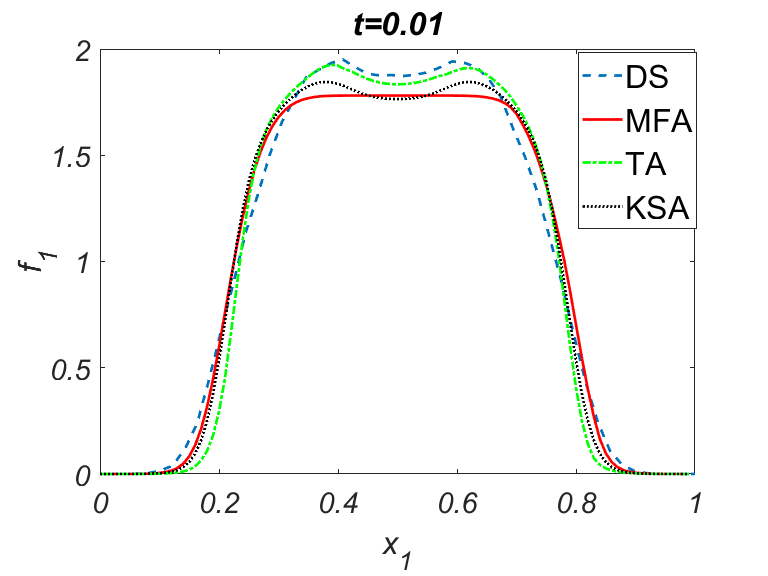}	
		\end{center}
		\caption{Approximations for $f_1(0.01,x)$ for the Morse interaction force with $c=0.2$, $D=0.045$, and initial conditions \eqref{tanIntCond}.  }
		\label{altmorsecomp}
	\end{figure}

\begin{figure}[!htb]
		\centering
		\includegraphics[width=.45\textwidth]{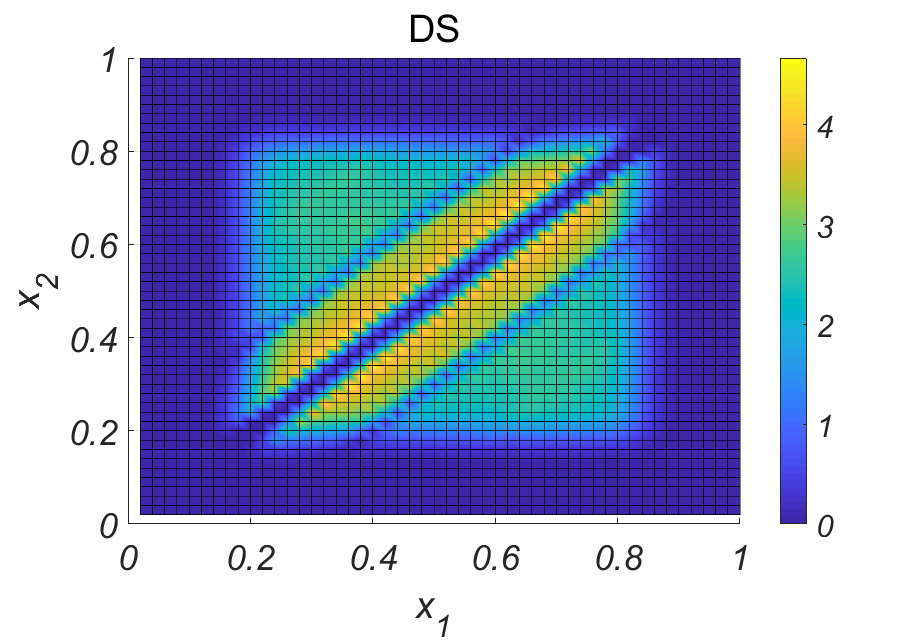}	
		\includegraphics[width=.45\textwidth]{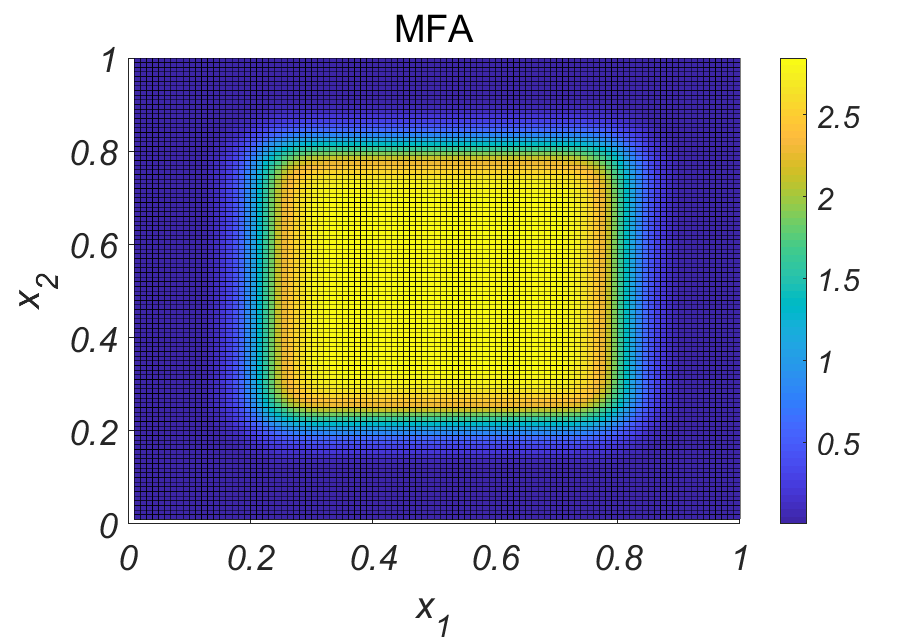}\\
		\includegraphics[width=.45\textwidth]{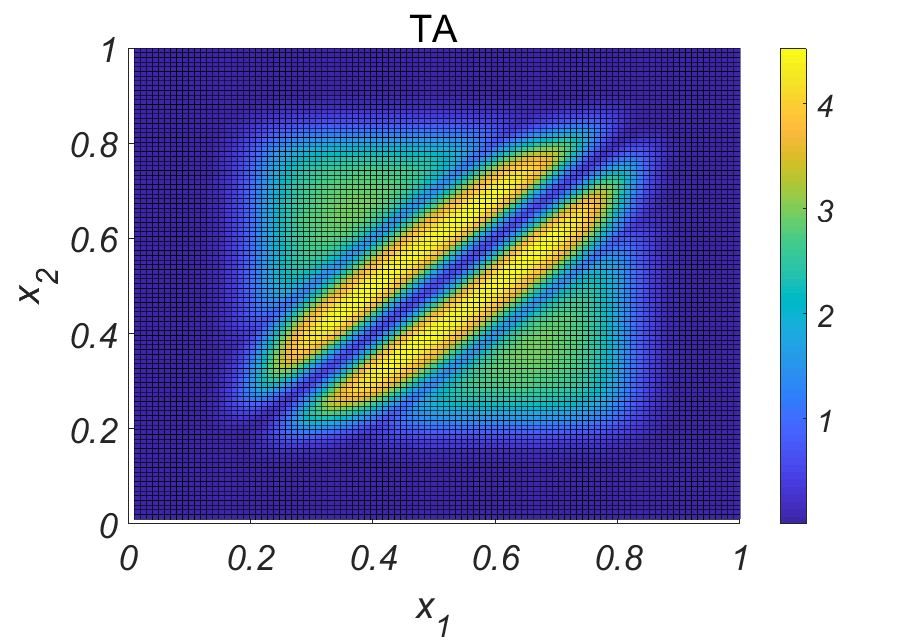}
		\includegraphics[width=.45\textwidth]{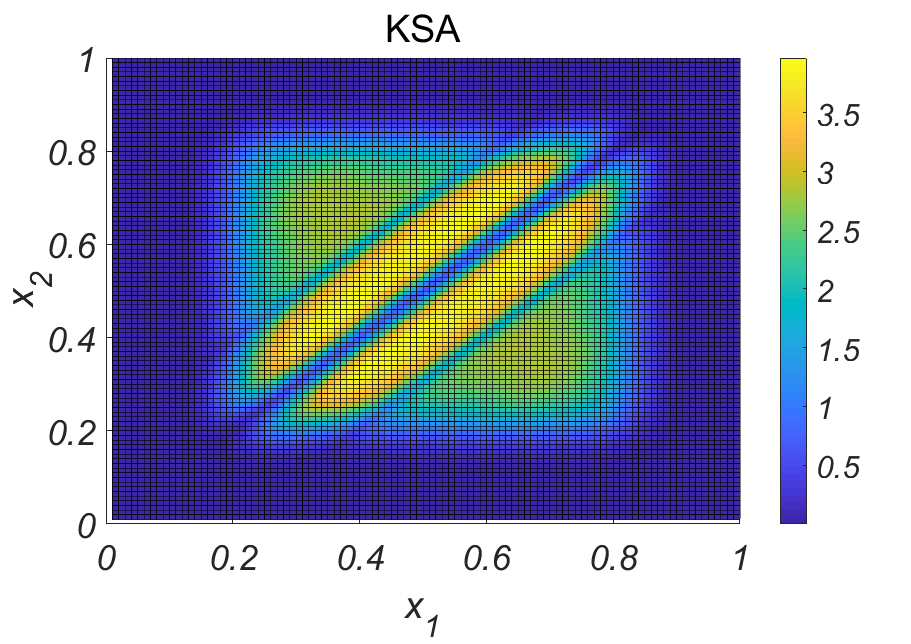}
		\caption{Approximations for $f_2$ at $t=0.01$ for the Morse interaction force with $c=0.2$.  Top left: Direct Simulations, top right: Mean Field Approximation, bottom left: Truncation Approximation, bottom right: Kirkwood Superposition Approximation. 	\label{Mfigf21}}
	\end{figure}

	First we consider the system  \eqref{Morse1}-\eqref{Morse2} with $N=5$, $D=0.045$ and $c=0.2$.
	 Two distinct peaks in the plot of $f_1$ obtained by direct simulations are observed, see fig.~\ref{altmorsecomp}.  Both TA and KSA  capture these peaks, and TA approximates the peaks more accurately. MFA does not exhibit any peaks.  
	   In capturing $f_2$, both TA and KSA capture the low values along the diagonal lines, $x_1=x_2$ and $x_1 \approx x_2 \pm 0.2$, whereas MFA does not, see fig.~\ref{Mfigf21}.  Additionally, TA captures the maximum values of $f_2$ better than KSA as well as the narrowness of the peaks' width.
	
\begin{figure}\centering
		\includegraphics[width=.5\textwidth]{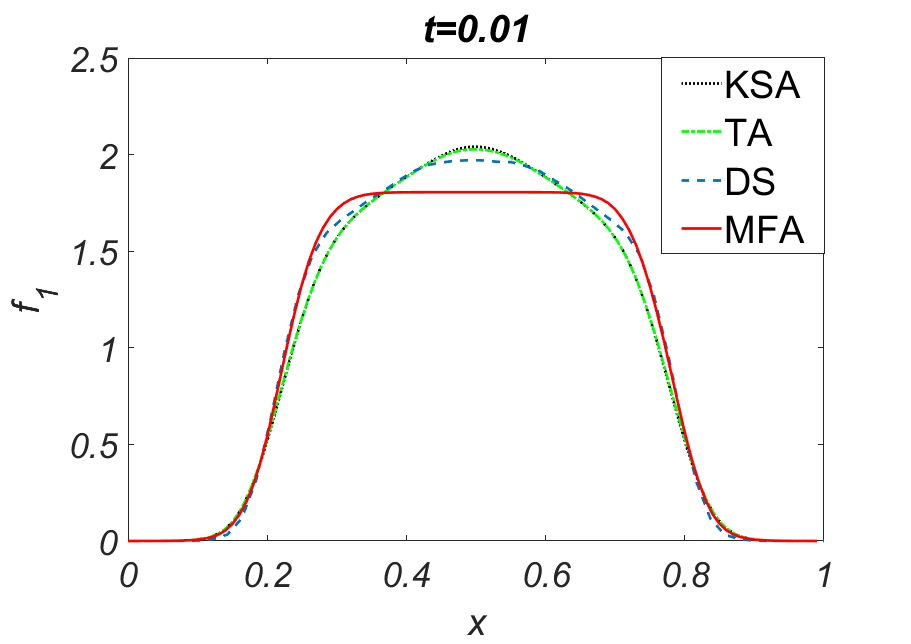}
		\caption{This figure shows the various approximations for $f_1(0.01,x)$ for the system \eqref{IBM} with the Morse interaction force with $c=0.3$ and $D=0.045$ and initial conditions \eqref{tanIntCond}.  }
		\label{M2figf1}
	\end{figure}

\begin{figure}[!htb]
		\centering
		\includegraphics[width=.45\textwidth]{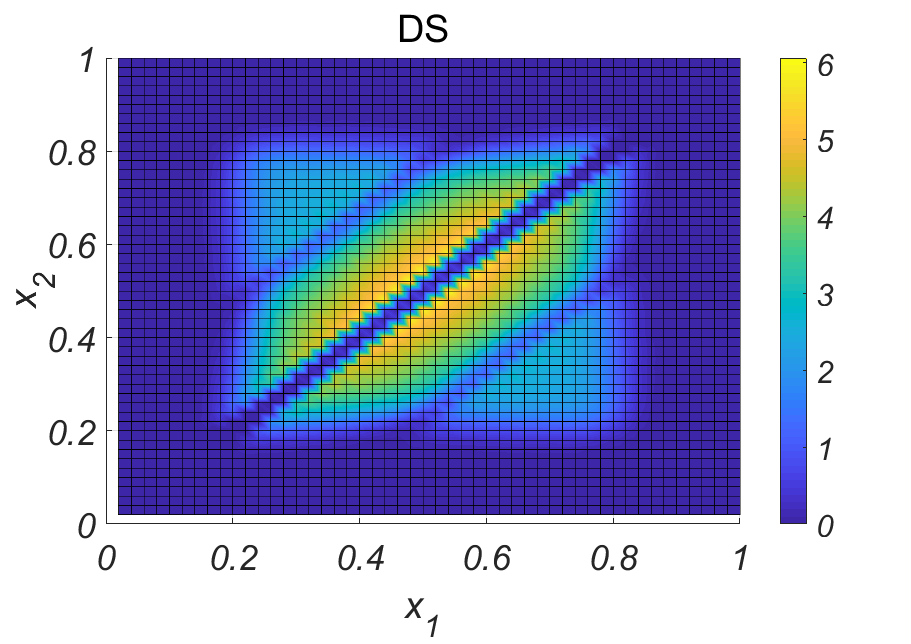}
		\includegraphics[width=.45\textwidth]{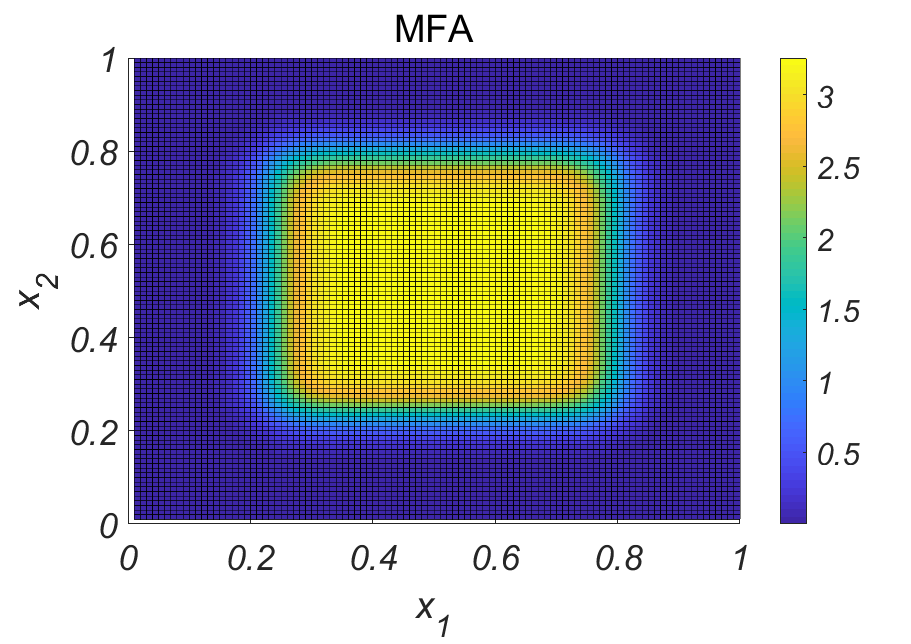} \\
		\includegraphics[width=.45\textwidth]{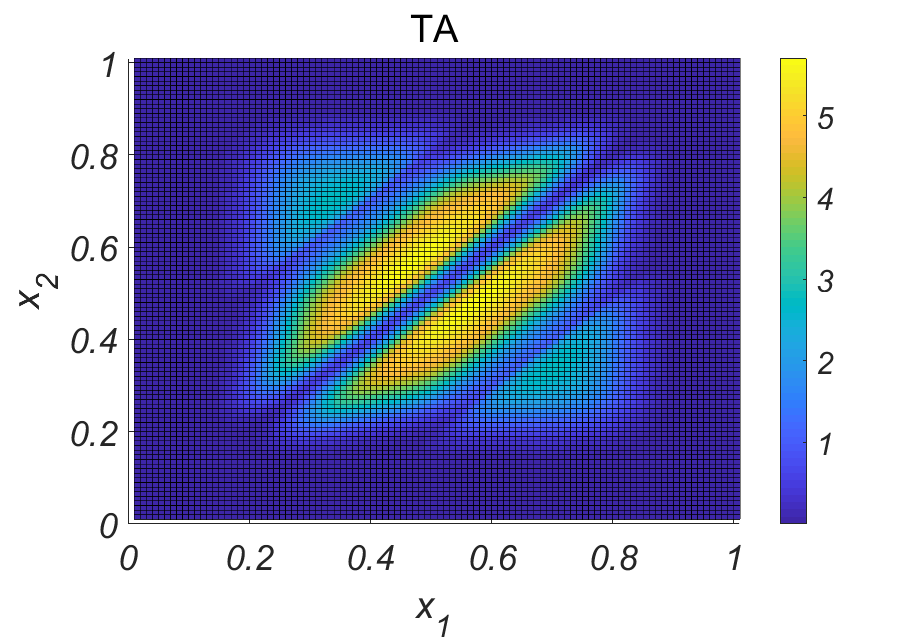}
		\includegraphics[width=.45\textwidth]{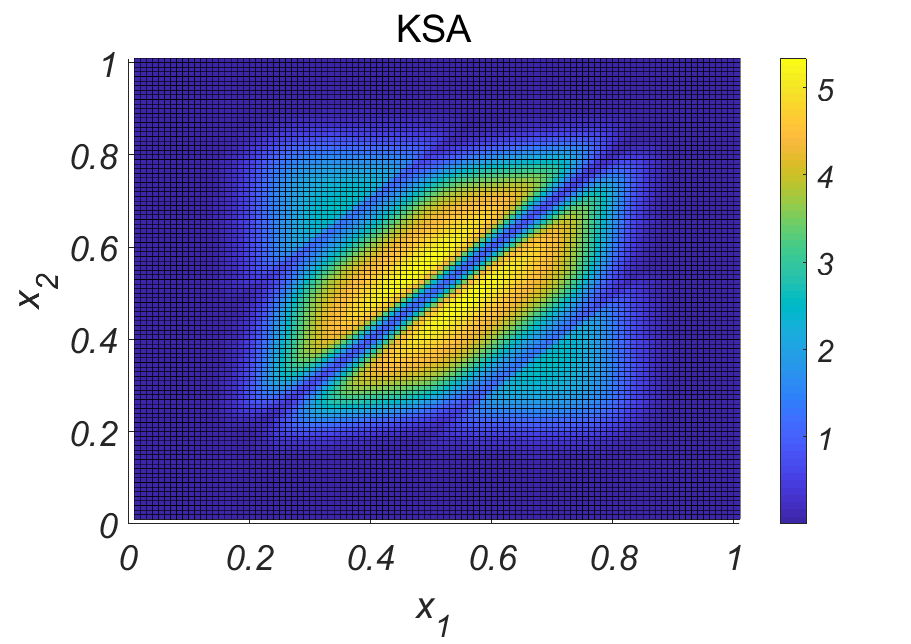}
		\caption{Approximations for $f_2$ at $t=0.01$ for the Morse interaction force with $c=0.3$.  Top left: Direct Simulations, top right: Mean Field Approximation, bottom left: Truncation Approximation, bottom right: Kirkwood Superposition Approximation. }
			\label{M2figf24}
	\end{figure}

	Next consider the system \eqref{Morse1}-\eqref{Morse2} with a larger range of interactions, specifically, $c=0.3$ with all other parameters remaining the same: $N=5$ and $D=0.045$.  Note that increasing the range of interactions effectively increases the strength of attraction between particles in the system.  In this case the one-particle probability distribution function $f_1$ has a single peak at center, see fig.~\ref{M2figf1}. TA and KSA both capture the peak, while MFA does not.  Comparing the approximations of $f_2$ depicted in fig.~\ref{M2figf24}, we see that both TA and KSA capture low  values along the diagonal lines, $x_1=x_2$ and $x_1 \approx x_2 \pm 0.3$, whereas MFA does not.  
	
	
	\subsection{Kuramoto interaction force}
	\label{subsec:kuramoto}
	
	Among all models used for the description of synchronization phenomena, the Kuramoto model is the most popular one and it was successfully used in various branches of science such as chemistry, physics, neural science, biology and even social science, \cite{Kuramoto1975,Pik2003,Ace2005}. In general, this model considers $N$ oscillators so that each oscillator has the phase $X_i(t)$ at time $t$, and the oscillators are coupled by the following attracting interaction force (which we call here the Kuramoto interaction force):     
	\begin{equation}\label{kuramoto_force}
	\hat u_{\text{K}}(x)=K\sin(2\pi x).
	\end{equation}
	The sub-index $\text{K}$ in the left hand side of \eqref{kuramoto_force} stands for ``Kuramoto" and the parameter $K=2.0$ in the right hand side of \eqref{kuramoto_force} is the strength of interactions.  
	
	A distinguishing feature of the Kuramoto model is that in addition to pairwise interactions, each oscillator also has a given intrinsic frequency $w_i$.  The resulting individual based system is 
	\begin{equation}
	\text{d}X_i(t)= w_i\text{d}t+\frac{1}{N} \sum_{j=1}^N \hat{u}_K(X_i-X_j) \text{d}t +\sqrt{2 D}\,\text{d}W_i(t),\quad \text{for $i=1,..,N$.} \label{IBM_w} 
	\end{equation}
	 The interactions are purely attractive, therefore the particles tend to occupy a single location at each moment of time moving with the same frequency/velocity.  However, if values of frequencies $w_i$ are high, then they dominate the attractive interactions and in this case the one-particle probability distribution function becomes uniform.
	

	  
	  In numerical simulations we choose $N=5$ and $D=0.045$. 	The frequencies $w_i$ are random variables, independently and identically distributed with the uniform distribution on $(-1,1)$.  
	 
	\begin{figure}
		\center{
			\includegraphics[width=.5\textwidth]{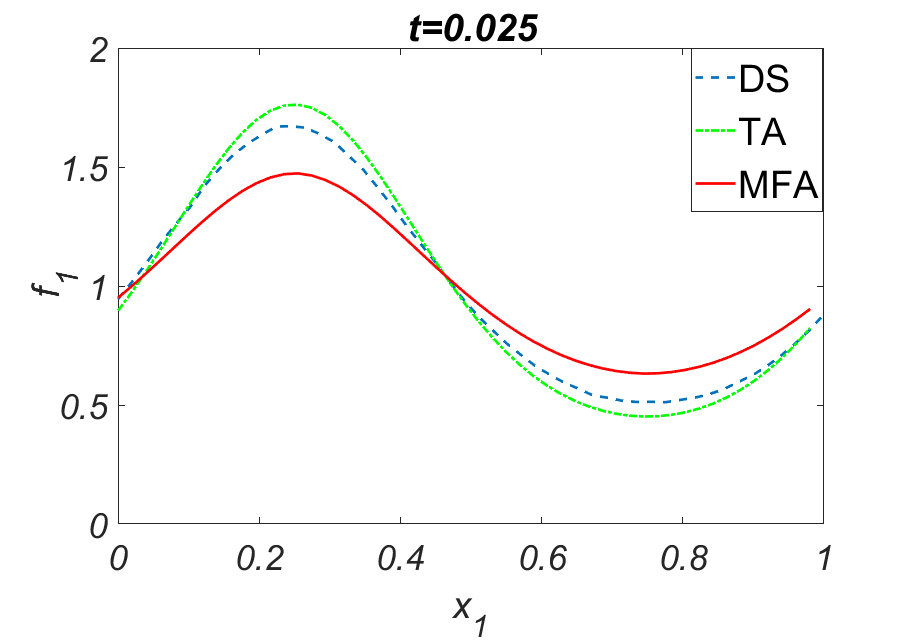} }
		\caption{This figure shows the various approximations for $f_1(0.025,x)$ for the system \eqref{IBM_w} with the Kuramoto interaction force and initial conditions given by\eqref{intcondsin}.  } \label{kura1}
	\end{figure}
	\begin{figure}[!htb]
		\centering
		\includegraphics[width=.45\textwidth]{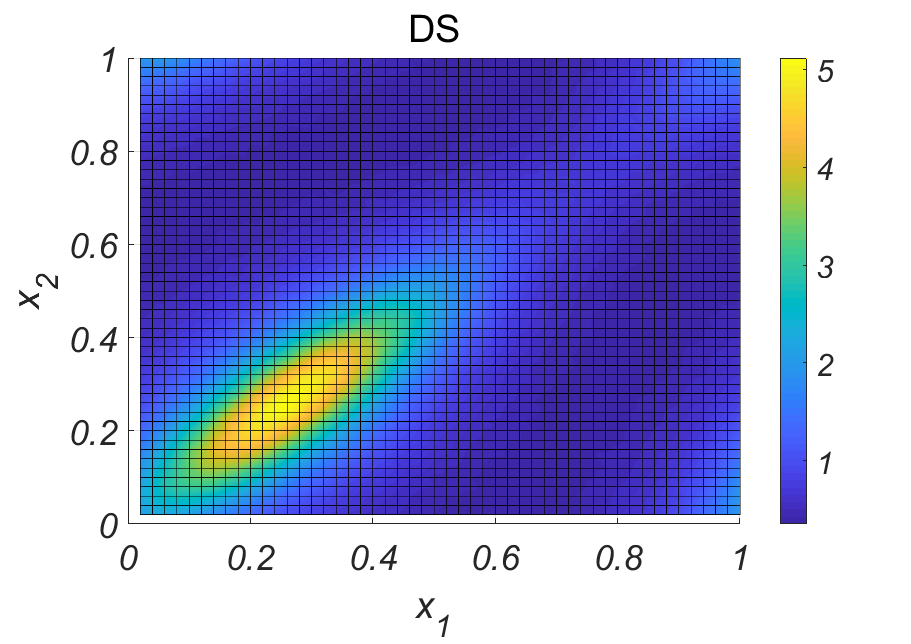}	\\
		\includegraphics[width=.45\textwidth]{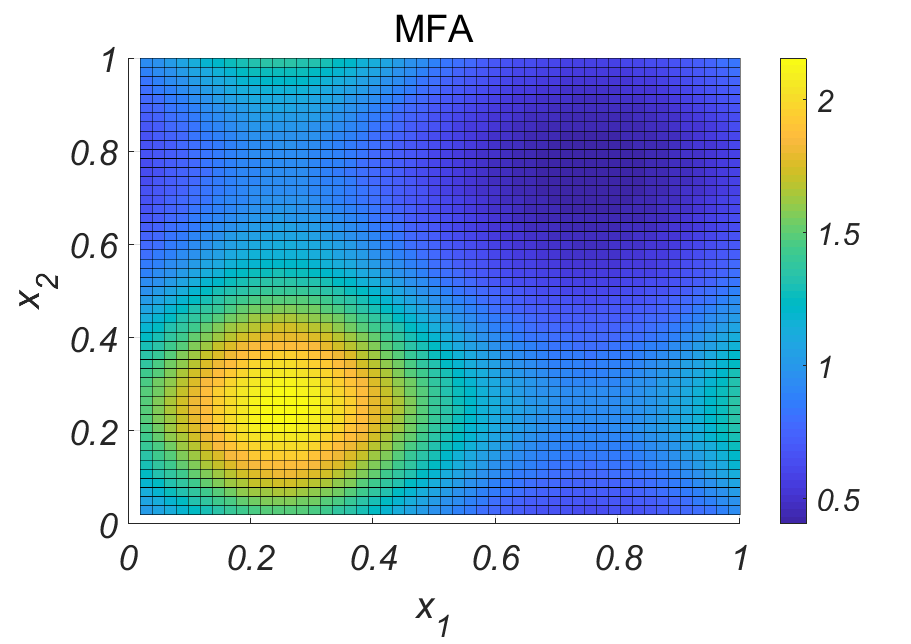}
		\includegraphics[width=.45\textwidth]{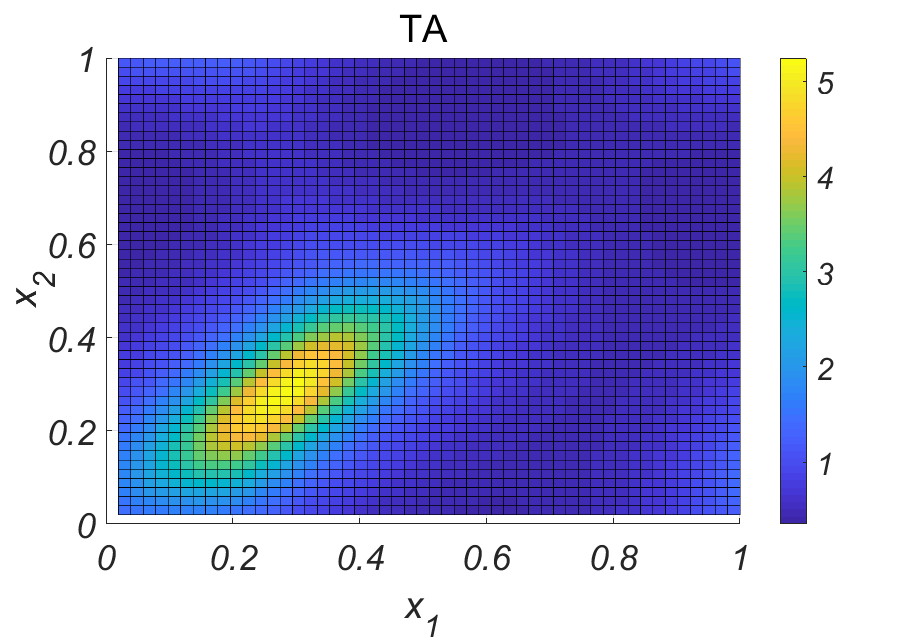}
		\caption{\label{Kfigf23} The figures show the various approximations for $f_2$ at $t=0.025$ with the Kuramoto interaction force as follows:  top: Direct Simulations, bottom left: Mean Field Approximation, bottom right: Truncation Approximation. }
	\end{figure}
	
	From fig. \ref{kura1} we see that both  TA and MFA exhibit  a peak in $f_1$ and TA is more accurate than MFA. TA also approximates direct simulations more accurately than MFA.  In comparing the approximations to direct simulations, TA captures the increase of $f_2$ near the diagonal $x_1=x_2$ whereas MFA does not. KSA was not used here since the introduction of intrinsic frequencies significantly increases computational complexity of KSA.
	
	\section{Conclusions}
	\label{sec:conclusion}
	In this paper three continuum approximations of a system of interacting particles $-$ Mean Field Approximation, Kirkwood Superposition Approximation, and Truncation Approximation $-$ were tested and compared to direct simulations of the individual based system for various types of interactions.   It was shown that in all the considered cases TA and KSA performed noticeably better than MFA.  When comparing TA and KSA, TA performed significantly better for all tested interactions except the repulsive interaction. For attractive interactions TA was stable while KSA was not.  The major advantage TA had over KSA was the computational complexity.  Due to the form of integral terms TA has significantly shorter computational time.  This advantage becomes more important as the dimension of a problem is increased.  In comparison to direct simulations,  continuum approximations are faster as they do not depend on the number of particles $N$ and do not require many realizations to take into account randomness of initial particles' locations.  
	
	\begin{acknowledgements}
	PEJ was partially supported by NSF Grant 1614537, and NSF Grant RNMS (Ki-Net) 1107444.  LB and MP were supported by NSF DMREF Grant DMS-1628411. 
	\end{acknowledgements}
	
	%
	

\end{document}